\documentclass[12pt,preprint]{aastex}
\begin{document}

\title {The Disk and Extraplanar Regions of NGC 2403 
\altaffilmark{1} \altaffilmark{2} \altaffilmark{3}}

\author{T. J. Davidge}

\affil{Herzberg Institute of Astrophysics,
\\National Research Council of Canada, 5071 West Saanich Road,
\\Victoria, B.C. Canada V9E 2E7\\ {\it email: tim.davidge@nrc.ca}}

\altaffiltext{1}{Based on observations obtained with the
MegaPrime/MegaCam, a joint project of the CFHT and CEA/DAPNIA, 
at the Canada-France-Hawaii Telescope (CFHT), which is operated by 
the National Research Council (NRC) of Canada, the Institut National des 
Sciences de l'Univers of the Centre National de la Recherche 
Scientifique (CNRS) of France, and the University of Hawaii}

\altaffiltext{2}{Visiting astronomer, Canada-France-Hawaii Telescope (CFHT), which is 
operated by the National Research Council (NRC) of Canada, the Institut National des 
Sciences de l'Univers of the Centre National de la Recherche 
Scientifique (CNRS) of France, and the University of Hawaii}

\altaffiltext{3}{This publication makes use of data products from the Two Micron 
All Sky Survey, which is a joint project of the University of
Massachusetts and the Infrared Processing and Analysis Center/California
Institute of Technology, funded by the National Aeronautics and Space 
Administration and the National Science Foundation.}

\begin{abstract}

	Wide field images obtained with WIRCam and MegaCam on the 
Canada-France-Hawaii Telescope are used to probe the spatial distribution 
of young stars in the Sc galaxy NGC 2403. Bright main sequence stars 
and blue supergiants are detected out to projected galactocentric distances of 
$\sim 14$ kpc ($\sim 7$ disk scale lengths) along the major axis. The star 
formation rate (SFR) in the disk of NGC 2403 during the past 10 Myr has been 
1 M$_{\odot}$ year$^{-1}$ based on the number of bright main sequence 
stars. The radially-averaged number density of red supergiants 
(RSGs) per unit $r-$band surface brightness is 
constant throughout the disk, indicating that (1) RSGs follow the integrated 
$r-$band light, and (2) the SFR per unit mass has been constant with 
radius when averaged over time scales of a few tens of millions of years. 
The mean color of RSGs varies with galactocentric distance, suggesting that 
there is a metallicity gradient among recently formed stars. A comparison of the 
radially-averaged number density of bright main sequence stars also 
indicates that the SFR per unit stellar mass in NGC 2403 has been $\sim 
3 \times$ higher than in NGC 247 during recent epochs, and this is in rough 
agreement with what would be predicted from the far infrared fluxes 
of these galaxies. Finally, the data are used to 
investigate the extraplanar regions of NGC 2403. A population of M giants 
with peak brightness M$_K = -8$ is detected at projected distances 
between 12 and 14 kpc above the disk plane, and six new globular 
cluster candidates are identified. 

\end{abstract}


\section{INTRODUCTION}

	Galaxies with disk-like morphologies appear to be common at high redshift 
(e.g. Marleau \& Simard 1998; Labbe et al. 2003), and these objects have almost certainly 
experienced significant evolution to the present day. It is likely that many of these 
systems have been consumed in the formation of larger galaxies, 
including spiral galaxies in the nearby Universe. While the frequency 
of merger activity that forms Milky-Way sized objects is predicted to have 
peaked $8 - 12$ Gyr in the past (Zentner \& Bullock 2003; Bullock \& Johnson 2005), 
significant merger activity likely continues to the present day. Indeed, 
based on the frequency of infrared luminous galaxies at intermediate redshifts, 
Hammer et al. (2005) argue that $\sim 75\%$ of intermediate mass spiral galaxies 
experienced a major merger in the past 8 Gyr. Simulations suggest that such large mergers 
may not destroy the disk of the dominant galaxy, but may simply provide 
material for disk growth (Volker \& Hernquist 2005; Robertson et al. 2006). 

	If disk assembly continues to recent epochs then evidence of this 
activity might be found in nearby spiral galaxies, including the large scale 
properties of the disks themselves. As time progresses it is 
likely that objects with progressively higher angular momentum will be accreted. Disks 
will then grow from the inside out, and there will be two observational consequences. 
First, the assimilation of material with high angular momentum 
will cause disks to evolve to larger sizes, and so distant disks will be physically 
smaller than nearby objects. Unfortunately, the measurement of disk size is not a 
straight-forward task. Narayan \& Jog (2003) point out that the measurement of disk size 
from surface brightness measurements is complicated by photometric depth effects, 
and argues that this has lead to the underestimation of disk sizes in previous 
studies. Our understanding of how to gauge disk size may also be uncertain, as 
deep photometric studies of moderately large samples of galaxies reveal a diversity in 
characteristic light profiles at large galactocentric radii, with some systems 
showing truncated disk light profiles, while others show no evidence of disk 
truncation, and even the onset of structures that are structurally 
distinct and more extended than the inner disks (Erwin, Beckman, \& Pohlen 2005). Still, 
Trujillo \& Aguerri (2004) and Trujillo \& Pohlen (2005) find that the characteristic 
scale lengths of disks near redshift $\sim 1$ are smaller than those of nearby 
systems, indicating evolution in disk size (but see also Lilly et al. 1998). 

	The second observational consequence of inside-out disk formation is 
that the stellar content of the outer and inner disks may differ. 
Gas accreted from satellites will almost certainly have 
experienced a chemical enrichment history that differs from that of the larger seed disk. 
Stars that form from material with only modest chemical enrichment will have a distinct 
chemical signature, and might produce a break in underlying abundance trends that were 
set in place earlier, although the radial mixing of disk material due to viscosity or 
tidal stirring will blur such trends. An age gradient may also be present if the outer 
regions of the disk have recently accreted gas-rich systems, which may cause the 
outer disk to have a younger mean age than the inner regions.

	The study of the distribution of stars 
perpendicular to the disk plane also provides clues into disk evolution. 
Deep imaging studies of the extraplanar regions of disk-dominated galaxies 
reveal a moderately metal-poor extraplanar component 
(Davidge 2003; Davidge 2005; Davidge 2006a; Seth, Dalcanton, \& de Jong 2005), 
the presence of which was orginally detected in integrated light studies 
(e.g. Dalcanton \& Bernstein 2002). The metallicity and spatial distribution of the 
extraplanar stars is suggestive of a disk origin, and Seth et al. (2005) interpret the 
distribution of stars in their sample as the result of heating by spiral density waves 
and/or scattering by molecular clouds, although the accretion of companion 
systems is another possible mechanism (e.g. Yoachim \& Dalcanton 2006; Bensby et 
al. 2005).

	As one of the closest ensembles of late-type galaxies, 
the M81 group is an important laboratory for probing disk evolution. 
The Sc galaxy NGC 2403 is an interesting 
member of this group as its integrated brightness, morphology, 
and environment do not differ greatly from those of 
the Local Group Sc galaxy M33, which is a benchmark for understanding 
late-type spiral galaxies. NGC 2403 has a modest entourage of four dwarf 
irregular galaxies and one dwarf spheroidal galaxy. The crossing time
of the NGC 2403 mini-system is $\sim 10$ Gyr; hence, the system is not yet virialized 
(Karachentsev et al. 2002). NGC 2403 is close enough that its brightest 
stars can easily be resolved with ground-based facilities, and it is 
viewed at an oblique angle, permitting the 
distribution of stars both on and off the disk plane to be investigated. 

	There have only been a few previous investigations of the stellar content of NGC 
2403, and these examined a modest fraction of the total galaxy. The innermost 
regions of NGC 2403 harbour a nucleus that is bluer than that in M33 (Davidge \& Courteau 
2002), suggesting that nuclear star formation occured more recently in NGC 2403. Still, 
when considering the properties of the brightest resolved 
red supergiants (RSGs) and asymptotic giant branch (AGB) stars near the nucleus, 
Davidge \& Courteau (2002) conclude that -- when averaged over 
Gyr timescales -- the star-forming histories of the central 
regions of NGC 2403 and M33 have been similar. 

	Davidge (2003) used deep $g'r'i'z'$ GMOS images to study the stellar 
content along the minor axis of NGC 2403. Projected distances between 5 and 11 kpc on 
the plane of the sky were covered, and a mix of disk and halo stars were found. 
Young stars in these data are restricted to projected 
galactocentric distances $< 10$ kpc along the disk plane. As for older stars, the mean 
color of the red giant branch (RGB) changes across the 
field. This is almost certainly a metallicity effect, such that [Fe/H] 
$\sim -0.8$ closest to the center of the galaxy and [Fe/H] $\sim -2.2$ 
in the outermost portion of the field. The latter metallicity is suggestive of a 
halo component with no disk contribution, while the former is consistent with a 
disk population. A number of bright AGB stars were also identified 
throughout the field, suggesting that an intermediate age population is present.

	With the advent of visible and near-infrared 
imagers that cover large angular fields it is now possible to survey the disk of 
NGC 2403 in its entirety, and conduct a comprehensive census of its brightest stars. 
In the current study, wide-field near-infrared and visible wavelength 
observations of NGC 2403 obtained with WIRCam and MegaCam on 
the Canada-France-Hawaii Telescope (CFHT) are used to chart the spatial 
distribution of stars at various stages of evolution. The goals of this 
study are (1) to investigate the spatial distribution of the brightest stars with 
respect to the integrated light, with particular emphasis on the outer regions of the 
disk, (2) to search for possible systematic stellar population gradients throughout the 
disk, (3) to further investigate the nature of the extraplanar AGB stars detected by 
Davidge (2003), and (4) to search for globular clusters.

	The distance modulus measured from Cepheids by 
Freedman \& Madore (1988) is adopted throughout this study, 
as is the foreground extinction from the Schlegel, Finkbeiner, 
\& Davis (1998) maps, which is A$_B = 0.172$. The correction for internal extinction 
in NGC 2403 is expected to be signficant, given that the galaxy 
is actively forming stars and is observed at an oblique angle. To account for this, 
the internal extinction computed by Pierce \& Tully (1992), as deduced from the model of 
Fouqu\'{e} \& Tully (1985), is applied to all CMDs and LFs in \S\S 4 and 5.

	The paper is structured as follows. Details of the observations and the 
data reduction procedures can be found in \S 2, while the photometric 
measurements and the CMDs are discussed in \S 3. The spatial distributions 
of disk stars at various evolutionary stages are examined in \S 4, while the 
properties of stars and candidate globular clusters in the extraplanar regions 
are explored in \S 5. A summary and discussion of the results follows in \S 6.

\section{OBSERVATIONS AND DATA REDUCTION}

\subsection{MegaCam Data}

	Deep $g', r'$, and $i'$ images of NGC 2403 were obtained 
with the CFHT MegaCam during the night of November 22, 2003. 
MegaCam is a mosaic of thirty six $2048 \times 4612$ pixel$^2$ CCDs that 
are deployed in a $4 \times 9$ configuration. There is a 13 arcsec gap between 
adjacent CCDs. Each pixel samples 0.19 arcsec on a side, and a 
$\sim 1$ degree$^2$ area is covered with each exposure. 
A more detailed description of MegaCam has been given by Boulade et al. (2003).

	The data were recorded as a series of 200 sec exposures with NGC 2403 
positioned near the center of the detector mosaic. Eight such 
exposures were recorded in $g'$, five in $r'$, and six in $i'$. 
The initial processing of the MegaCam data was done with the Elixir pipeline 
at CFHT, and this consisted of bias subtraction and flat-fielding. A calibration 
frame to suppress interference fringes was also subtracted from the $i'$ data.

	Additional processing was done at HIA Victoria. The 
Elixir-processed images were spatially registered and then median-combined. The 
results were trimmed, such that each CCD samples the area that is common to 
all exposures. Stars in the processed data have FWHMs of 1.2 arcsec in $g'$, 
1.1 arcsec in $r'$, and 1.0 arcsec in $i'$. The final $i'$ images of the ten 
MegaCam CCDs that sample the disk and inner halo of NGC 2403 are displayed 
in Figure 1.
 
\subsection{WIRCam Data}

	Moderately deep $J, H$, and $K'$ images of NGC 2403 were recorded 
with the CFHT WIRCam during the nights of November 19 and 20, 2005.
The detector in WIRCam is a mosaic of four $2048 \times 2048$ pixel$^2$ HgCdTe arrays 
that are deployed in a $2 \times 2$ layout. There is a 45 arcsec 
gap between adjacent arrays. Each pixel subtends 0.3 arcsec on a side, and 
a $20 \times 20$ arcmin$^2$ field is covered in a single exposure. 
Puget et al. (2004) give a more detailed description of WIRCam. 

	The data were recorded as a series of short exposures 
distributed over a six point dither pattern. The total exposure times were 2640 sec 
in $J$, 1680 sec in $H$, and 2400 sec in $K$. The data were processed with a 
standard pipeline for near-infrared imaging that includes 
(1) dark subtraction, (2) flat-fielding, (3) the removal of 
interference fringes and thermal emission artifacts from warm objects along the light 
path, and (4) the subtraction of a DC sky level from each image. 
The calibration frames used in the third step were constructed from 
observations of a background field. The processed images were spatially 
registered, median-combined, and then trimmed to the area common to all exposures.
Stars in the processed data have FWHMs of 0.9 arcsec in $J$, and 0.8 arcsec in $H$ 
and $K'$. The region covered by the WIRCam data is marked in Figure 1.
 
\subsection{CFHT Adaptive Optics Data}

	Deep $J, H,$ and $K'$ images were obtained of a field in the disk of NGC 2403 
with the CFHT Adaptive Optics Bonnette (AOB) and KIR imager on the night of March 9, 
2001. The AOB is a natural guide star system, and a 
detailed description is given by Rigaut et al. (1998). The disk 
field is centered on the star GSC 04120--00764, 
which was the reference source for AO correction; the star is 
labelled in Figure 1. The sky conditions were not photometric,
as thin clouds were present when these data were obtained.

	The KIR camera contains a $1024 \times 1024$ pixel HgCdTe array with 0.034 
arcsec pixels; therefore, a $35 \times 35$ arcsec $^2$ field is covered with each 
exposure. The data were recorded as a series of 60 second exposures, with a total 
exposure time of 1200 sec filter$^{-1}$. The data were processed with the 
same pipeline for near-infrared imaging that was described in \S 2.2.
The seeing was poor when these data were recorded, and the image quality 
of the final data ranges from 0.2 arcsec in $K'$ to 0.4 arcsec in $J$. 
Still, the angular resolution of these data is superior to that obtained with 
WIRCAM, and so they can be used to assess the effects 
of crowding at the faint end of the WIRCam observations. 

	Observations of the center of NGC 2403, which were obtained with the AOB and were
used previously by Davidge \& Courteau (2002), are also considered in this study. These 
data have an angular resolution of 0.15 -- 0.20 arcsec FWHM, depending 
on the filter. Additional details of the observations and their reduction 
can be found in Davidge \& Courteau (2002).

\section{RESULTS}

\subsection{Photometric Measurements}

	The photometric measurements were made with the point spread function 
(PSF) fitting routine ALLSTAR (Stetson \& Harris 1988). The target lists, 
preliminary brightnesses, and PSFs that serve as the input to ALLSTAR were 
obtained from routines in the DAOPHOT (Stetson 1987) package. 
Photometry of sources in the WIRCam and MegaCam data was done on an array-by-array 
basis. While this is more labour intensive than measuring brightnesses from a single 
mosaiced image, as a single PSF must be constructed for 
each array, this process better accounts for variations in image quality and 
optical distortions across the (wide) science fields of both instruments.

	It is standard procedure for a Landolt (1992) standard star field to be 
observed at the beginning of every clear night of MegaCam 
queue observing. The standard star observations from various nights are then 
used to establish a single zeropoint for each run, and the results are 
inserted into image headers during Elixor processing. These 
zeropoints were used to calibrate the MegaCam observations of NGC 2403.

	The WIRCam observations were calibrated using objects in the 2MASS Point 
Source Catalogue (Cutri et al. 2003). Only sources with $K < 15$ were considered, 
as the uncertainties in the $K-$band 2MASS photometry are smaller than $\pm 0.1$ 
magnitudes in this brightness range. The uncertainty in the zeropoints are typically 
$\pm 0.02 - 0.03$ magnitudes filter$^{-1}$ array$^{-1}$. The 
individual arrays have measurably different zeropoints that 
differ by $\pm 0.05 - 0.08$ magnitude, depending on the filter.

	The CFHT AOB observations were calibrated using the brightnesses 
of GSC 04120 -- 00764 in the 2MASS Point Source Catalogue; no 
other sources in the AOB field are bright enough to serve as an independent 
calibrator. The use of a single object to define the photometric calibration introduces 
obvious uncertainties. However, it is demonstrated in \S 3.4 that the 
brightnesses and colors of stellar sequences in the AOB data are broadly consistent 
with those in other datasets. Anisoplanicity, which causes the shape of the PSF to 
vary with distance from the AO guide star, is another potential source of 
uncertainty in photometric measurements made from AO-corrected data. 
However, previous experience with the AOB indicates that anisoplanicity is typically not 
significant over the field imaged with KIR, even in data recorded when 
atmospheric conditions are such that low Strehl ratios are delivered (e.g. Davidge 2001).

	Artificial star experiments were run to assess completeness and estimate 
the scatter in the photometry due to photon noise and crowding. 
The artificial stars were assigned brightnesses and colors that are typical of objects 
in NGC 2403. The results depend on the distance from the 
center of the galaxy, although the 50\% completeness limit throughout most of 
the NGC 2403 disk occurs near $r' = 23.5$ and $K = 19.5$. The predicted 
$1-\sigma$ color dispersions at $r' = 23.5$ are typically $\sigma_{g'-r'} = \pm 0.14$ 
magnitudes and $\sigma_{r'-i'} = \pm 0.12$ magnitudes throughout most of the disk, 
while for the WIRCAM data the predicted color dispersions at $K = 19.5$ are typically 
$\sigma_{J-K} = \pm 0.19$ magnitudes and $\sigma_{H-K} = \pm 0.25$ magnitudes.

	Background galaxies are a significant source of contamination near the faint 
limit of these data, and efforts to identify extended objects will be beneficial. 
The number of background galaxies can be estimated from published 
surveys. McCracken et al. (2000) summarize $K-$band galaxy 
counts from a number of studies, and the data in their Figure 1 indicates that 
there are $\sim 2.3 \times 10^4$ galaxies deg$^{-2}$ when $K < 19.5$. This 
implies that $\sim 2600$ galaxies with $K < 19.5$ might be present in the WIRCam 
science field. The veracity of this estimate is affected by cosmic 
dispersion, and the GalaxyCount (Ellis \& Bland-Hawthorn 2007) routine suggests 
that there is a $\pm 5 - 10\%$ uncertainty in the number of galaxies.

	In an effort to reject obviously extended objects from the MegaCam and WIRCam 
data, a threshold value of the DAOPHOT {\it sharp} parameter, which measures the 
compactness of a source, was computed from the artificial star experiments. The 
ability to reject extended objects at the faint end is complicated by (1) noise, which 
hinders the calculation of reliable shape statistics 
for individual objects, and (2) image quality, which smears the signal 
for small objects. Still, the application of even a relatively conservative threshold 
allows objects that have obviously non-stellar light profiles to be culled from the data.

\subsection{The MegaCam Data}

	The $(r',g'-r')$ and $(i', r'-i')$ CMDs of sources in four 
selected MegaCam CCDs are shown in Figure 2. The CMDs of these particular CCDs 
are shown because they illustrate the colors and brightnesses of key features 
that will be used throughout this study. Data from a larger number of 
CCDs, which sample the entire disk of NGC 2403 and its surroundings, are considered 
in \S 4.

	The left hand panels of Figure 2 show the CMDs of control regions that are 
distinct from the disk of NGC 2403. The bright end of the control region $(r', g'-r')$ 
CMD is dominated by the blue and red foreground star sequences that are predicted by 
Galactic structure models (Fan 1999). The blue sequence is made up 
of stars near the main sequence turn off (MSTO) in the Galactic halo, while the 
red sequence is made up of lower mass main sequence stars in the Galactic disk.
The blue sequence in the MegaCam data is less well defined than the 
red sequence, in qualitative agreement with the Fan (1999) simulations.
The North Galactic Pole (NGP) model considered by Fan (1999) predicts a 
characteristic color $(g'-r') \sim 0.3$ for the blue sequence at $r' = 20$, 
while for the red sequence $(g'-r') \sim 1.3$. NGC 2403 is not near the NGP; still, 
when $r' = 20$ the blue foreground sequence in the control field CMD has 
$(g'-r') \sim 0.4$, while for the red sequence $(g'-r') \sim 1.3$.

	The right hand panels of Figure 2 show the CMDs of stars in 
the disk of NGC 2403. There is a broad blue plume 
in both the $(r', g'-r')$ and $(i', r'-i')$ CMDs that is made up of blue 
supergiants (BSGs) and high mass main sequence stars. This sequence 
falls to the left of the blue foreground star plume, thereby simplifying the task of 
detecting BSGs and massive main sequence stars in regions of NGC 2403 that have 
low stellar density.

	Red supergiants (RSGs) dominate the red portions of the 
NGC 2403 $(r', g'-r')$ and $(i', r'-i')$ CMDs. The brightest 
RSGs form a finger-like sequence in the $(i', r'-i')$ CMD, that peaks 
near $i' \sim 20$, and has $r'-i' \sim 0.5$. Stars evolving on the AGB 
are found in the $(i', r'-i')$ CMDs when $i' > 23$ and $r'-i' 
\geq 1$; the faint limit of the $g'$ data is such that the 
brightest AGB stars are not sampled in the $(r', g'-r')$ CMDs. 

\subsection{The WIRCam Data}

	The $(K, J-K)$ CMDs of sources in two quadrants of the WIRCam 
array that sample the major and minor axes of NGC 2403 are 
shown in the top row of Figure 3; the $(K, H-K)$ CMDs are not shown as they are 
very similar to the $(K, J-K)$ CMDs. The CMDs are dominated by stars in the Galactic disk 
when $K < 16$, and foreground stars account for the vast majority of objects 
with $J-K \leq 1$. Low mass stars in the Galactic disk populate the sequence with $J-K 
\sim 0.8$, the color of which is indicative of M dwarfs (Bessell \& Brett 1988).

	Contamination from background galaxies is significant at faint magnitudes. This 
being said, the observed near-infrared spectral-energy distributions (SEDs) of 
background objects differ significantly from those of the majority of stars 
in NGC 2403. The mean redshift of galaxies with $K \sim 17$ is $<z> \sim 
0.4$ (Cowie et al. 1996). Applying the k-corrections from Mannucci et al. (2001), a 
normal $z = 0.4$ galaxy will have $H-K \sim 0.8$ and $J-K \sim 1.6 - 1.7$; in fact, a 
clump of objects with $J-K \sim 1.7$ and $K < 18$ can be 
seen in the $(K, J-K)$ minor axis CMD. For comparison, M giants have 
$H-K \sim 0.2 - 0.3$ and $J-K \sim 1 - 1.3$ (Bessell \& Brett 1988). 
The comparatively red colors of intermediate redshift galaxies are a consequence 
of the first overtone CO bands, which are amongst the strongest absorption features 
in the spectra of normal galaxies, being red shifted out of the bandpass of 
the $K$ filter. Even though intermediate redshift galaxies have 
infrared SEDs that differ from the majority of evolved stars in NGC 2403, it was decided 
not to cull the WIRCam data by applying a color threshold to avoid deleting the reddest 
stars in NGC 2403. Still, the reader should keep in mind that the majority of the 
reddest objects are probably background galaxies.

	The objects detected in the WIRCam data that belong to NGC 2403 
are RSGs and AGB stars. RSGs form a plume in the $(K, J-K)$ CMD 
that is slightly redder than the foreground star sequence, while 
objects that are evolving on the upper AGB form a tangle of red stars with $K \geq 18$ 
and $J-K < 2$. The RSG sequence in NGC 2403 peaks near M$_K \sim -11$, 
while the AGB sequence peaks near $-9.5$. These peak brightnesses are similar to 
those seen in other nearby spiral galaxies (e.g. Davidge 2005; 2006a,b).

	The MegaCam and WIRCam data have been combined to create a data set 
with broad wavelength coverage, and therefore increased sensitivity to metallicity 
and age. The WIRCam and MegaCam data have different pixel sampling, and this 
could introduce systematic effects in the photometry, 
especially in crowded fields. In an effort to reduce such systematic 
effects, the images from the $r'$ MegaCam CCDs that overlap with the WIRCam 
data were re-sampled to the 0.3 arcsec pixel spacing of the WIRCAM data, and 
DAOPHOT was run on the result. Whereas combining the $g'$ and $K$ data 
would offer even greater wavelength leverage, the number 
of objects that are common to both the $g'$ and $K$ observations 
is modest, and it was decided that combining the $r'$ and $K$ data would serve 
as a suitable compromise between wavelength coverage and sample size. 

	The $(K, r'-K)$ CMDs constructed from these data are shown in the bottom 
row of Figure 3. The number of objects in these CMDs is relatively modest when 
compared with the WIRCam or MegaCam datasets alone since (1) 
the gaps between the individual detectors in the WIRCam and MegaCam 
arrays do not overlap, with the result that sky coverage is reduced, and 
(2) only the brightest and most highly evolved stars in NGC 2403 
are sampled when observations spanning such a wide wavelength range are combined. 
Indeed, the main sequence, which is prominent in the $(r', g'-r')$ and $(i', r'-i')$ 
CMDs, is abscent in the lower panels of Figure 3
because of the faint limit of the $K-$band observations; consequently, the 
spray of objects with $r'-K < 2$ is made up of foreground stars.

	RSGs and AGB stars dominate the $(K, r'-K)$ CMD when $K > 16$; the majority of 
RSGs have $r'-K < 4$, whereas the AGB stars have $r'-K > 3$. The red cut-off in 
the lower panels of Figure 3 is defined by the faint limit of the $r'$ observations, and 
the intrinsically reddest evolved stars in the NGC 2403 disk are probably 
missing from these CMDs. Indeed, the reddest stars in the $(K, r'-K)$ CMDs have colors 
that suggest that the latest M giants are not sampled. 
Fukugita et al. (1996) give a transformation equation relating 
$r'$ and $V$, and this can be used to compute $V-K$ colors 
if a relation between $B-V$ and $V-K$ is also adopted. 
For the current work it is assumed that $B-V \sim 0.33 \times (V-K)$, based 
on the properties of solar neighborhood M giants (Bessell \& Brett 1988). 
The reddest objects in the $(K, r'-K)$ CMDs are then found to 
have $V-K \sim 7$, which is appropriate for giants of spectral type M6 (Bessell \& 
Brett 1988). For comparison, giants with later spectral-types are expected in NGC 2403
(\S 4.1), and it is likely that these stars fall below the faint limit of the $r'$ data. 

\subsection{$AOB+KIR$ Data}

	The $(K, H-K)$ and $(K, J-K)$ CMDs obtained from the AOB $+$ KIR images are 
shown in the upper row of Figure 4. The stars in the CMDs are evolving on the AGB; the 
absence of bright RSGs in these data is a consequence of the small area sampled with KIR, 
coupled with the modest spatial density of these objects in this portion of NGC 2403. 
It should be emphasized that the peak AGB brightness in these data is highly uncertain, 
not only because the photometric calibration is based on only one object (\S 3.1), 
but also because of photometric variability and stochastic effects. Still, 
the peak brightness in the upper row of Figure 4 is $K \sim 18$, which 
corresponds to M$_K \sim -9.5$, and is consistent with the AGB-tip brightness 
predicted by models (Girardi et al. 2002).

	The AOB disk field falls in a gap between WIRCam arrays, and so it 
is not possible to make a direct comparison with the WIRCam photometry. However, the 
central field observed by Davidge \& Courteau (2002) is in the WIRCam images; thus, 
a direct comparison can be made between the AOB$+$KIR photometry from 
Davidge \& Courteau (2002) and the WIRCam photometry. This is a rigorous test of the 
WIRCam photometry, as the central $\sim 30$ arcsec of NGC 2403 is a crowded environment. 
Unlike the disk AO field, the central AO field was 
observed during photometric conditions. The photometric calibration of these data is 
based on observations of UKIRT faint standards (Hawarden et al. 2001), 
and the uncertainty in the photometric zeropoints is $\pm 0.01$ magnitudes. 

	The $(K, H-K)$ and $(K, J-K)$ CMDs of the central field from Davidge \& Courteau 
(2002) are shown in the middle row of Figure 4, while the CMDs of the same region 
extracted from the WIRCam data are shown in the 
bottom row. The peak brightnesses in the AOB and WIRCam central field data sets 
are in excellent agreement. While there is more scatter in the WIRCam data, 
the mean colors of the sequences in the two data sets are also in reasonable agreement. 
An object-by-object comparison of the $K-$band brightnesses of sources with $K$ between 
16 and 17.5 gives a difference $\Delta K = -0.09 \pm 0.09$ magnitudes, where 
the difference is in the sense WIRCam -- AOB and 
the error is the standard uncertainty in the mean. 
It thus appears that crowding does not greatly bias the photometric 
properties of the brightest stars in the WIRCam observations, even near the center 
of NGC 2403. This is because the AGB and RSG sequences on near-infrared CMDs are 
more extended along the magnitude axis than at visible wavelengths, leading to a 
greater degree of contrast with respect to the main body of fainter stars. 

\section{THE SPATIAL DISTRIBUTION OF STARS IN THE DISK OF NGC 2403}

\subsection{General Trends and Comparisons with Models}

	The CMDs of stars in 2 kpc-wide annular intervals are shown in Figures 5 - 
8. The computation of R$_{GC}$ for each star assumes that the sources are in the 
disk plane, which in turn is inclined to the line of sight by 53$^o$ 
(Pierce \& Tully 1992). The $(K, H-K)$ CMDs are not shown because they 
provide information that is similar to the $(K, J-K)$ CMDs. 

	The CMDs of the $0 - 2$ kpc and $2 - 4$ kpc intervals do not go as deep as 
those at larger R$_{GC}$ owing to the higher stellar density near the center of 
NGC 2403. This being said, the inner annuli sample relatively small areas on the sky, 
and so there is only modest contamination from foreground stars and background galaxies. 
The foreground star sequence becomes more pronounced in the CMDs as R$_{GC}$ increases 
as progressively larger areas on the sky are sampled.

	The plume of main sequence stars and BSGs that is a prominent feature in the 
$(r', g'-r')$ CMDs has been used to probe interstellar 
extinction associated with young stars throughout the disk of NGC 2403. To do this, 
the mean color and width of the main sequence in the $(r', g'-r')$ CMDs 
were computed for stars with $r' = 22 \pm 0.25$. The dispersion about the mean 
color for these stars is remarkably constant with radius, with $\sigma_{g'-r'} = \pm 
0.2$ magnitudes. This is more than a factor of two larger than the width 
expected solely from observational errors. In addition to differential reddening, 
it is likely that stellar evolution, binarity, and photometric variability 
also contribute to broadening the main sequence; nevertheless, adopting the plausible 
assumption that these other sources of broadening are fixed throughout the NGC 
2403 disk, then the detection of a uniform main sequence width suggests that the 
amount of differential reddening does not change throughout the disk of NGC 2403.

	The mean color of the blue sequence varies with radius in NGC 2403, with $<g'-r'> 
= 0.01$ between 0 and 4 kpc, and $<g'-r'> = -0.1$ between 6 and 10 kpc. 
This change in mean color suggests that there is a radial gradient in the mean 
extinction associated with young stars in the disk of NGC 2403, such that $E(g'-r')$ 
in the central 4 kpc is 0.1 magnitudes higher than at larger radii. 
Using the model of Tully \& Fouqu\'{e} (1985), Pierce \& Tully (1998) 
estimate that the internal extinction in NGC 2403 is A$_B \sim 0.20$, 
which corresponds to $E(g'-r') \sim 0.05$ using the relations in Appendix C of Schlegel, 
Finkbeiner, \& Davis (1998). This internal extinction is adopted for R$_{GC} > 4$ 
kpc throughout this study, while $E(g'-r') = 0.15$ is adopted for R$_{GC} < 4$ 
kpc. It should be kept in mind that the higher internal reddening found when R$_{GC} 
< 4$ kpc is based on only the youngest stars in the disk, some of which are almost 
certainly still embedded in natal dust and gas, and that the internal extinction for 
older populations may be lower. 

	Young stars are detected out to R$_{GC} > 12$ kpc in the NGC 
2403 disk. Indeed, the main sequence can be seen in the $(r', g'-r')$ CMD of 
the $12 - 14$ kpc interval, while a population of objects with 
main sequence-like colors is also seen in the $14 - 16$ kpc interval. 
The brightest blue objects in the 12 - 14 kpc interval have 
$r' \sim 22$, which corresponds to M$_V \sim -6$. This is the 
approximate peak brightness expected for main sequence stars (Humphreys \& McElroy 1984), 
and so it appears that star formation is on-going in the outer regions of the NGC 
2403 disk.

	RSGs form a distinct finger that peaks near $i' \sim 20$ 
in the $(i', r'-i')$ CMDs of stars with R$_{GC}$ 
between 6 and 10 kpc. Models that trace stellar evolution 
predict that the color of the RSG sequence at a fixed age is sensitive to metallicity. 
In Figure 9 the $(i', r'-i')$ CMDs in the 2 -- 4 and 8 -- 10 kpc intervals are 
compared with Z = 0.008 and Z = 0.019 isochrones from Girardi et al. (2004). 
The locus of the RSG sequence in the outer interval is roughly matched by the 
Z = 0.008 models, and a similar metallicity was estimated for RSGs at intermediate radii 
in NGC 247. The comparisons in Figure 9 further suggest that the stars that populate 
the RSG plume have log(t) between 7.0 and 7.5. 

	The RSG sequence in the $2 - 4$ kpc 
interval is broader than at larger radii, and extends to redder colors. 
Models with metallicities between Z = 0.008 and Z = 0.019 roughly bracket the 
range of RSG colors in the $2 - 4$ kpc interval. The broad range in RSG 
colors is likely not due to patches of locally high internal reddening, given that 
the width of the main sequence in this interval is comparable to that at larger 
radii. Rather, the comparisons in Figure 9 suggest that RSGs in NGC 2403 have a similar 
metallicity when $r > 6$ kpc, but that at smaller radii the RSGs span a range 
of metallicities. 

	How do the metallicities estimated for RSGs in NGC 2403 
compare with abundance estimates for HII regions? Garnett 
et al. (1997) investigate the chemical compositions of HII regions throughout the 
disk of NGC 2403, and find radial gradients in the abundances of various 
elements. Consider those HII regions in the inner disk of NGC 2403. 
The sample of HII regions observed by Garnett et al. (1997) with R$_{GC} < 
4$ kpc typically have log[O/H] $\sim -3.3$, which is within $\sim 0.2$ dex of 
solar. There is also a dispersion in oxygen abundances in this region, with two 
HII regions having [O/H] $\sim 0.3$ dex lower than average. Based on these observations 
one might expect to find (1) a population of RSGs with moderately high metallicities, 
and (2) RSGs spanning a range of metallicities. These expectations are consistent 
with the range in metallicities of RSGs with R$_{GC} < 6$ kpc inferred from Figure 9. 

	Keeping in mind that [O/Fe] depends on star-forming history, 
there is reasonable agreement between the metallicities of RSGs and HII regions 
at larger R$_{GC}$ as well. At distances in excess of 
6 kpc from the center of NGC 2403 the HII 
regions tend to have log[O/H] $\sim -3.7$, or roughly one quarter solar. 
While based on only one point, located at R$_{GC} \sim 9.5$ kpc, Figure 6 
of Garnett et al. (1997) is consistent with a constant [O/H] when R$_{GC} > 6$ kpc. 

	The $(M_K, J-K)$ CMDs of stars in the $2 - 4$ kpc and $6 - 8$ kpc intervals are 
compared with Z = 0.008 and Z = 0.019 models from Girardi et al. (2002) in Figure 
10. The $J-K$ color of RSGs is less sensitive to metallicity than is $r'-i'$, and 
the Z = 0.008 and Z = 0.019 sequences have characteristic $J-K$ 
colors that differ by only $\sim 0.1$ magnitude. 
This has an impact on the width of the RSG sequence; unlike what is 
seen in the $(i', r'-i')$ CMD, the RSG plume in the $2 - 4$ kpc interval in Figure 
10 is comparatively well-defined, peaking near $K \sim 16$. The RSG plume 
is bracketed by the Z = 0.008 and Z = 0.019 models, and is matched 
best by models with ages $\geq 10$ Myr, in agreement with what is seen 
in the $(i', r'-i')$ CMDs. While RSGs are present in the $6 - 8$ kpc interval, 
the detection and characterization of individual RSGs at these R$_{GC}$ in the infrared 
is complicated by contamination from foreground stars, the majority of which have 
$J-K$ colors that are similar to RSGs. 

	The comparisons in Figure 10 underline the 
importance of observations at wavelengths longward of $\sim 1\mu$m 
for probing the AGB content of galaxies. Whereas AGB-tip stars with ages of at least 
1 Gyr are detected in the $(K, J-K)$ CMDs, the comparisons in Figure 9 indicate 
that only the most luminous AGB stars, which have ages that are considerably 
less than 1 Gyr, are detected in the $(i', r'-i')$ CMDs. It should also be recalled that 
the comparisons discussed in \S 3.4 indicate that the infrared photometric properties 
of bright AGB stars are not affected greatly by crowding.

	The $(M_K, r'-K)$ CMDs of the $2 - 4$ and $6 - 8$ kpc intervals are compared 
with Z = 0.008 and Z = 0.019 isochrones from Girardi et al. (2002) in 
Figure 11. Girardi et al. (2002) do not tabulate $r'$ brightnesses, and so these 
were computed from the $B$ and $V-$band entries using the transformation 
relation from Fukugita et al. (1996). An important caveat is that this transformation 
relation was calculated from blue objects, and so the transformation becomes 
progressively more uncertain towards redder colors.

	The RSG sequence in the $(M_K, r'-K)$ CMD of stars in the $2 - 4$ kpc interval 
is not well-defined, as might be expected given the range of RSG colors 
in the $(M_{i'}, r'-i')$ CMD of the same radial interval. Indeed, as in Figure 9 
the Z = 0.019 models in Figure 11 match the $r'-K$ colors of the reddest objects, 
while the Z = 0.008 models match the blue envelope of RSGs in this radial interval. 
The RSG sequence is better defined in the CMD of objects in the $6-8$ kpc 
interval. The RSG plume has $r'-K \sim 3$, and the red locus of the 
Z = 0.008 models is better able to match the RSG sequence than the Z = 0.019 models. 
In summary, the $(M_K, r'-K)$ CMDs support the conclusions 
obtained from the $(M_{i'}, r'-i')$ CMDs regarding the metallicities of the 
brightest RSGs in the disk of NGC 2403.

\subsection{Radial Trends and the Specific Frequencies of Stars in the Disk}

	Physical processes during disk formation and 
subsequent disk evolution may cause population gradients that will affect 
the properties of stars on spatial scales spanning much of the 
disk. Evidence for such trends have already been discussed; 
in \S 4.1 it was shown that the RSGs in the inner 
regions of NGC 2403 have a range of metallicities, whereas RSGs at larger radii have 
Z = 0.008. This suggests that the interstellar material from 
which these objects formed does not have a uniform metallicity; rather, there 
is metal-rich gas at small R$_{GC}$ that is not present at large R$_{GC}$.

	In addition to metallicity, mean age might also be 
expected to vary in the disk of NGC 2403. For example, 
if the central regions of the galaxy have been stoked with gas during recent epochs, then 
there might be an elevated density of young stars per unit mass in this portion of the 
galaxy. If high angular momentum gas has been accreted into the outer 
disk then there might be a higher density of young stars at large R$_{GC}$. 
To investigate the spatial distribution of objects in NGC 2403, two 
samples of stars with differing ages were identified using their locations 
on CMDs. Main sequence stars and BSGs were identified from 
the $(r', g'-r')$ CMDs, while RSGs were selected using the $(i', r'-i')$ 
CMDs. The portions of the CMDs used to identify these stars are 
indicated in Figure 12, which shows the CMDs of stars in the $6 -8$ kpc interval 
for illustrative purposes. 

	The spatial distributions of the blue and red stars are examined 
in Figure 13, where the projected (top row) and de-projected (lower row) 
locations of stars in each sample are compared. The blue stars trace the spiral 
structure of NGC 2403, and pockets of star-forming activity are seen. 
In contrast, the RSGs are more uniformly distributed; while a hint of spiral structure 
is evident, the knots of star-forming activity that are seen in the blue star 
distribution are absent. The relative behaviour of the 
two stellar types is consistent with stars being scattered from their 
place of birth as they age; similar behaviour has been documented 
in other galaxies (e.g. Thuan \& Izotov 2005).

	The comparisons in Figure 13 illustrate that while 
blue stars provide insight into the most recent episodes of star formation, 
as tracers of stellar content they may be prone to stochastic effects, especially 
in the outer regions of the disk. This is less of a concern for RSGs.
Before proceeding to a more quantifiable means of assessing the distributions of 
the blue and red stars, it should be noted that the space velocities of RSGs in 
NGC 2403 can be estimated from the comparison in Figure 13 knowing (1) their 
approximate ages, and (2) the amount of spatial filtering that is required to suppress 
the appearance of distinct star-forming regions. Comparisons with stellar isochrones 
suggest that the main body of RSGs have ages $\sim 50$ million years. As for the second 
quantity, this can be estimated by applying gaussian filters to blue images of NGC 2403, 
and visually assessing when individual knots in the spiral arms can no longer be 
distinguished. It was found that smoothing with a gaussian kernel with a 
standard deviation $\pm 0.8$ kpc suppresses individual star-forming regions, and 
produces a diffuse disk like that seen in the right hand panel of Figure 13. Hence, the 
velocity dispersion of bright RSGs in the disk of NGC 2403 is $\sim 10$ km 
sec$^{-1}$. While this is only an approximate quantity, with estimated 
uncertainties of $\pm 50\%$, it is similar to the velocity dispersion of 
solar neighborhood Cepheids (Wielen 1974).

	The light profile of NGC 2403 has been investigated at red wavelengths by 
Kent (1987), and in the remainder of this section we investigate the extent to which 
the radial distributions of RSGs and main sequence stars follow 
his light profile. In particular, the spatial distribution of young 
stars is traced by computing their specific frequencies. The specific 
frequency is a quantity that was originally developed by Harris \& van den Bergh (1981) 
to measure the number of globular clusters per unit galaxy brightness. Here, 
this basic concept is extended to stars by calculating the number of stars in a given 
evolutionary stage per unit magnitude interval per unit integrated brightness. 
The specific frequency of main sequence stars or RSGs will change with radius in NGC 2403 
if star-forming activity during the past few tens of millions of years has been 
concentrated in a given radial interval. This statistic is not sensitive to isolated 
pockets of star-forming activity, unless the localized SFR was of such intensity that 
the average annular SFR was significantly elevated. 

	The specific frequency measurements are based on the luminosity functions 
(LFs) of the blue and red stars within the color boundaries defined in Figure 12. 
The results, in which the number counts have been scaled as if each annulus samples 
a system with M$_r = -15$ using the $r-$band photometry from Kent (1987), 
are shown in Figures 14 and 15. The number counts have been 
corrected for light lost due to the gaps between detectors, as well as 
for contamination from foreground and background objects using 
the number counts of sources in the same color intervals 
in fields at large galactocentric distances along the minor axis of NGC 2403. The 
$r-$band surface brightness measurements used to compute the specific frequencies 
of stars with R$_{GC} > 6$ kpc are those that are marked as `extended by hand' 
in Kent's Table II. The dashed line in each panel is the mean LF of 
stars located between 4 and 12 kpc from the galaxy center; this mean LF is shown 
to provide a benchmark to aid in determining if gradients are present. The 
error bars show Poisson uncertainties. The LFs of stars in the $0 - 2$ kpc 
interval are not shown, as crowding restricts the faint limit of these data, 
while the data for the $12 - 14$ kpc and $14 - 16$ kpc intervals 
have been combined to boost the signal-to-noise ratio. 

	The specific frequency measurements in Figure 14 suggest that 
the youngest stars in NGC 2403 are not uniformly distributed with radius. While there 
is reasonable agreement in the number density of blue stars with $r' > 22$ and 
R$_{GC}$ between 4 and 12 kpc, the specific frequency of blue stars in this magnitude 
range in the $2 - 4$ kpc interval exceeds that at intermediate galactocentric radii. 
There are also departures from the mean trend in the $2 - 4$ kpc interval when $r' 
> 23$, although this is due to incompleteness at the faint end in the high density 
regions at small galactocentric radii. 

	The modest number of blue stars at large radii 
may affect the ability to search for population gradients in the 
outer disk. In fact, there is a net deficiency of stars with $r' < 22.5$ 
in the $12 - 16$ kpc interval in Figure 14, and this could arise if star formation 
in the outer disk has been more sporadic than at smaller radii. 
Such stochastic effects in star formation events are suppressed when 
longer time intervals are considered, such as is the case at the faint end of the LFs in 
Figure 14. It is thus significant that the specific frequency of blue stars at the 
faint end of the LF of the $12 - 16$ kpc interval matches that at intermediate radii.

	The specific frequency of RSGs, as defined using the boundaries indicated 
in the right hand panel of Figure 12, is investigated in Figure 15. 
The specific frequency of RSGs varies when $i' < 20$, although this may simply be due 
to flucuations in the numbers of the brightest and rarest RSGs. Indeed, the number 
density of RSGs with $i' > 20$ does not change throughout much of the 
disk of NGC 2403. Whereas there is evidence for an elevated 
specific frequency of blue stars in the $2 - 4$ kpc interval in Figure 14, the specific 
frequency of RSGs in this same interval is consistent with that at larger radii. 

	The RSGs used to construct the LFs in Figure 15 sample a larger range of 
ages than the BSGs and main sequence stars used to construct Figure 14. 
One consequence is that short timescale variations in the SFR should have a 
smaller impact on the LFs in Figure 15 than on the LFs in Figure 14. This is seen in 
the specific frequency measurements of RSGs in the $12 - 
16$ kpc interval, which agree with the mean LF at all 
brightnesses; the gaps in star counts that are seen near the bright end of the 
$12 - 16$ kpc LF in Figure 14, are largely absent in Figure 15. The comparisons in 
Figure 15 indicate that RSGs in the disk of NGC 2403 follow the $r-$band light profile 
measured by Kent (1987). Thus, when averaged over time scales of a few tens of 
millions of years, the radially-averaged SFR throughout the disk of NGC 2403 has been 
uniformly distributed with respect to stellar mass in the disk.

\section{THE EXTRAPLANAR REGIONS}

\subsection{AGB Stars in the Extraplanar Regions of NGC 2403}

	While the disk of NGC 2403 harbors the youngest, intrinsically brightest, 
stars in the galaxy, the extraplanar regions also contain stars that 
are bright enough to be detected with modest exposure times from ground-based telescopes. 
Davidge (2003) detected sources along the minor axis of NGC 2403
that are significantly brighter than the RGB-tip, and some of these are probably 
stars evolving on the AGB. The WIRCam data are of interest for identifying and 
characterizing such a component, as the contrast 
between the brightest AGB stars and any fainter underlying objects 
is greater in the near-infrared than at wavelengths shortward of $1\mu$m (\S 4.1). 
Moreover, the large field of view covered by WIRCam allows both legs of the 
minor axis (i.e. to the north east and south west of the galaxy) to be sampled 
with a single pointing.

	The stellar content in two sub-fields of the WIRCam data that have identical 
projected areas on the sky and sample the north east and south west portions of the 
minor axis of NGC 2403, are examined to determine if a moderately bright AGB population 
is present. Stars in the outer regions of the NGC 2403 disk may be a significant 
source of contamination along the minor axis, and in \S 4 it was demonstrated that young 
disk stars can be traced out to $R_{GC} \sim 16$ kpc. In an effort to minimize 
contamination from stars in the outer disk, the minor axis fields considered 
in this paper are restricted to regions where R$_{GC} > 16$ kpc along 
the disk plane, which corresponds to distances along the minor axis $\geq 8.5$ kpc. 
The fields, which will be referred to as NE and SW throughout the remainder of the 
paper, sample projected distances along the minor axis 
between 8.5 and 12.7 kpc from the disk plane. The NE 
field overlaps with the region investigated by Davidge (2003). These fields sample 
distances off the disk plane that exceed those probed by Seth et al (2005) in their 
sample of edge-on disk galaxies, but are similar to those examined by 
Mouhcine et al. (2005) in their study of nearby spirals. 

	The $(K, J-K)$ CMDs of the extraplanar fields are shown in Figure 16, and the 
CMDs of the NE and SW fields are very similar. The right hand panel of Figure 16 shows 
the CMD of a control field to the south and west of the SW field, which samples 
projected distances along the minor axis between 12.7 and 14.8 kpc and has 
an area that is comparable to that of each extraplanar field. The CMD of the control 
field is very similar to that of the NE and SW fields, despite sampling different 
distances off of the disk. It is evident that the vast majority of objects in the NE and 
SW fields are foreground stars or background galaxies. 

	In \S 3 it was argued that the majority of objects at the faint end of the 
WIRCam data are normal galaxies at intermediate redshift, which 
have $J-K > 1.6$, and so are redder than M giants in NGC 2403. 
It is then significant that the CMDs of the NE and SW fields contain a population 
of objects with $J-K < 1.6$ and $K > 19.5$ that is not present in the control field. 
The NE and SW fields contain comparable numbers of objects in this 
brightness and color interval, with 13 in the NE field, and 9 in the 
SW field; for comparison, only 3 objects are seen in the corresponding portion of 
the control field, and all of these are huddled close to $J-K 
= 1.6$. When $K > 19.5$ the number of objects with $J-K < 1.6$ in 
the NE and SW fields agree with that in the control field, indicating 
that the extraplanar fields contain a detectable number of objects only when M$_K > -8$; 
while stars with M$_K < -8$ may be present in the extraplanar regions of NGC 2403, they 
do not occur in large enough numbers to allow them to be identified from number counts 
alone.

	The RGB-tip in old systems has M$_K > -7$ (Ferraro et al. 2000), and 
so the stars detected in the NE and SW fields are evolving on the AGB. 
A crude age can be estimated from their peak brightness. 
Davidge (2003) measured the colors of bright RGB stars 
in the extraplanar regions of NGC 2403 and found that 
[Fe/H] $< -1.4$ outside of the disk. Assuming that the AGB stars 
have a similar low metallicity, then the isochrones tabulated by Girardi et al. (2002) 
predict that AGB stars with M$_K \sim -8$ have an age $\sim 3$ Gyr. There 
are significant uncertainties in such an age estimate, and these are discussed in \S 6.3.

\subsection{Globular Clusters}

	The WIRCam data can be used to characterize globular clusters in NGC 2403. 
Indeed, if the globular cluster LF (GCLF) of NGC 2403 is like that in M31 (e.g. Barmby, 
Huchra, \& Brodie 2001), then it will peak near M$_K \sim -10$, which corresponds to 
$K \sim 17.5$ at the assumed distance of NGC 2403. It can also be anticipated from the 
M31 GCLF that most of the globular clusters in NGC 2403 will have $K$ between 
15.5 and 18.5, and so are well above the faint limit of the WIRCam data.
In addition. all but the most compact globular clusters in NGC 2403 will 
appear as obvious non-stellar objects in the WIRCam data. 
Galactic globular clusters that belong to the old halo typically have half light radii 
$\sim 2 - 3$ parsecs, while young halo clusters may have even larger sizes (Mackey 
\& van den Bergh 2005). Stars in the WIRCam data have FWHM $\sim 0.8 - 0.9$ arcsec, 
which corresponds to $\sim 12 - 14$ parsecs at the distance of NGC 2403. Hence, 
the majority of globular clusters in NGC 2403 will tend to have angular sizes that 
are at least a few tenths of an arcsec larger than stars in the WIRCam images. 

	Battistini et al. (1984) used photographic images to search 
for clusters in NGC 2403, and found 19 candidates. 
Five of the clusters marked in Figure 1 of Battistini et al. (1984) have been identified 
in the WIRCam data, and the near-infrared brightnesses and colors of these are listed in 
Table 1. These objects all have non-stellar light profiles, and the approximate 
characteristic size of each object, after accounting for the seeing disk, is listed in 
the last column of Table 1. The characteristic sizes of these objects tend to be 
larger than those of old halo Milky-Way globular clusters, and this is probably due to 
two reasons. First, in addition to {\it bona fide} globular clusters, 
the Battistini et al. (1984) sample also contains open clusters, stellar 
associations, and background galaxies. Second, while the image quality 
of the data used by Battistini et al. (1984) was not specified, it is probably 
poorer than that obtained with WIRCam, and this will skew the identification of cluster 
candidates to objects with larger sizes.

	Also listed in Table 1 is a $V-K$ color computed from the $r'-K$ and 
$B-V$ colors using Equation 23 of Fukugita et al. (1996). The Battistini et al. 
cluster candidates have a range of $V-K$ and $J-K$ colors. The majority of 
globular clusters in NGC 2403 would be expected to have $J-K < 1$ if they are like the 
old clusters in the M31 cluster system. Three of the cluster 
candidates in Table 1 (F1, F46, and F16) have $J-K$ colors that are consistent with this 
prediction. However, two cluster candidates (F14 and F19) have the very red $J-K$ and 
$V-K$ colors that are characteristic of background galaxies (\S 3.3), rather than 
of globular clusters. Radial velocity measurements will provide a 
sure means of determining if F14 and F19 are galaxies at intermediate redshift, 
or clusters in NGC 2403 with large AGB populations, and hence red colors.

	The WIRCam images were searched for additional globular cluster candidates 
using three screening criteria: (1) a non-stellar light profile, (2) 
$J-K < 1$, and (3) a location outside of the main body of the disk. 
Each of these criteria introduces biases. The first criterion 
biases against the detection of the smallest, most compact, clusters, although 
some of the candidates found here have characteristic sizes that are smaller than 
the peak of the size distribution of old Galactic halo clusters (e.g. Figure 4 
of MacKey \& van den Bergh 2005). The candidates were also inspected by eye to identify 
obvious non-cluster objects, such as the bulges of spiral galaxies. 
The second criterion introduces a bias against clusters that 
may contain a significant AGB population. AGB stars dominate the light output for systems 
with ages near $\sim 1$ Gyr (e.g. Maraston 1998), and clusters near this age will have 
very red $J-K$ colors. However, the absence of very luminous AGB-tip stars in the 
extraplanar regions suggests that such objects are likely not present in the 
fields we have surveyed. While clusters with ages $\sim 1$ Gyr may not be expected 
outside of the disk, the discovery of clusters with ages of $5 - 10$ Gyr may not be 
surprising, as some of the globular clusters in M33 appear not to be as 
old as those in the Milky-Way (e.g. Sarajedini et al. 1998).
Finally, the third criterion, which was introduced to 
prevent contamination by old open clusters and asterisms, means that the 
portion of the galaxy where the density of globular clusters is expected to be 
greatest was not searched.

	The locations and photometric properties of the six cluster candidates 
found in the WIRCam data are listed in Table 2. The co-ordinates are reliable to 
$\pm 1 - 2$ arcsec. If assumed to be at the distance of NGC 2403 then 
these objects have characteristic radii of a few parsecs, and so 
are comparable in size to the majority of Milky-Way globular clusters.

	Combining the six objects found in the 
WIRCam data with the four clusters with $J-K < 1$ in the Battistini et al. (1984) sample, 
then only $\sim 10$ likely cluster candidates have been identified so far 
in NGC 2403, and a number of others remain to be discovered. 
The number of globular clusters that might belong to NGC 2403 can be estimated by 
assuming that the cluster system is like that in M33, which is morphologically similar 
to NGC 2403. Christian \& Schommer (1982) found 13 globular 
clusters in M33; the survey was not complete, and more clusters are likely present. 
Given that the M$_K$ of NGC 2403 is $\sim 0.9$ magnitudes brighter than M33 (Garrett 
et al. 2003), then the NGC 2403 cluster system should contain at least $\sim 30$ 
globular clusters. The majority of these will likely be at 
small R$_{GC}$, where the cluster density is highest, and so will be 
projected against (or seen through) the disk. A problem searching for clusters in the 
disk is that there may be contamination from old disk clusters. 
There is also the potential for stellar asterisms to 
masquerade as young or intermediate age globular clusters, as has been found in M31 
(Cohen, Matthews, \& Cameron 2005). The secure detection of globular clusters in this 
environment will thus require images with high angular resolution.

\section{SUMMARY AND DISCUSSION}

	Images obtained with WIRCam and MegaCam on the CFHT have been used to 
investigate the bright stellar content in the nearby Sc galaxy NGC 2403. 
The science fields of both instruments sample the entire disk 
and much of the halo of the galaxy in a single pointing. The result is an unprecedented 
census of the brightest main sequence stars, BSGs, RSGs, and AGB stars that spans 
the $0.5 - 2.5\mu$m wavelength interval. A new sample of globular cluster candidates 
is also identified. The data are used to search for large-scale radial trends in 
the spatial distribution of the youngest stars, with the goal of charting radial 
trends in the star-forming history during recent epochs, and determining the extent 
of the young stellar disk. 

\subsection{Star Formation in the Disk of NGC 2403 and Comparisons with NGC 247}

	The intensity of star-forming activity per unit stellar disk mass 
during the past 10 Myr appears to have varied with radius 
in NGC 2403. Evidence for this comes from the specific frequency measurements 
of bright main sequence stars $+$ BSGs in the $2 - 4$ kpc interval, which are higher 
than at larger radii. For comparison, the specific frequencies 
of RSGs, which sample a range of ages that exceeds that covered by the brightest main 
sequence stars, indicate that these objects tend to be uniformly distributed with 
older stars throughout the disk of NGC 2403. 
Thus, any increase in the SFR in the $2 - 4$ kpc interval must have occured on a 
timescale shorter than the onset of the RSG phase of evolution, which corresponds 
to $\sim 10$ Myr. 

	The recent galaxy-wide SFR in NGC 2403 can be estimated from the number of bright 
main sequence stars. To compute the recent SFR the mean specific frequency 
measurements of blue stars with $r'$ between 20 and 22.5 in Figure 14 are used; 
while there is some radial variation in the mean specific frequency of these 
objects in the central few kpc, these are minor for the purposes of the computation 
of a global SFR. The Z = 0.008 models from Girardi et al. (2002) indicate that main 
sequence stars with masses between 19 ($r' = 22.5$) and 50 ($r' = 20$) M$_{\odot}$ are 
sampled in this brightness interval. If it is assumed that all the stars in this 
brightness interval are evolving on the main sequence, then 
there are 5460 main sequence stars in NGC 2403 within this mass range. 
The Kroupa, Tout, \& Gilmore (1993) mass function predicts that 
stars in this mass range account for 2.5\% of all stars that form. 
A SFR during the past $\sim 10$ Myr of $\sim 1$ M$_{\odot}$ year$^{-1}$ then results.

	How does the SFR computed from main sequence stars compare with that calculated 
from other indicators? The SFR computed from the far-infrared (FIR) flux measured by 
Condon et al. (1996) is 0.13 M$_{\odot}$ year$^{-1}$ using the Condon (1992) calibration. 
Dale et al. (2005) show that the ratio of 70$\mu$m and 160$\mu$m fluxes are 
correlated with SFR in galaxies. Combining the calibration in their Figure 15 with the 
fluxes given in their Table 1 predicts a mean star-forming density of 
$\sim 0.01 - 0.03$ M$_{\odot}$ year$^{-1}$ kpc$^{-2}$ in NGC 2403, depending on the 
calibrating galaxy. With an effective aperture radius of 5.8 kpc (Garrett et al. 2003), 
then the predicted SFR in the disk of NGC 2403 based on 70 and 160$\mu$m fluxes 
is $\sim 1 - 3$ M$_{\odot}$ year$^{-1}$, which agrees with that obtained above from 
main sequence star counts.

	Davidge (2006a) probed the stellar content of NGC 247 with MegaCam. 
A direct comparison between the stellar contents of NGC 247 and NGC 2403 
is of interest as these galaxies have similar morphologies and distances.
In fact, the radial photometric properties of RSGs in NGC 247 and NGC 2403 show broad 
similarities. The dispersion in $r'-i'$ colors of bright RSGs suggests that stars with 
metallicities as high as solar are present in the central 6 kpc of 
NGC 2403, while at larger radii the RSGs have Z = 0.008. A similar trend is seen 
in the $(i', r'-i')$ CMDs of NGC 247 in Figure 2 of Davidge (2006a). 
It thus appears that stars that formed recently near 
the centers of both galaxies did so from gas that had a range in chemical 
abundances, with the most chemically enriched material having a metallicity 
that was almost twice that at larger radii. Both of these results are consistent with 
the properties of HII regions in NGC 2403 (Garnett et al. 1997).

	The specific frequencies of stars in NGC 247 and NGC 2403 can be used to 
compare the recent star-forming histories of these galaxies in a differential manner. 
Main sequence stars $+$ BSGs and RSGs in the radial intervals 2 -- 6 kpc 
were identified in NGC 247 using the criteria defined in \S 4.2, and 
in Figure 17 the specific frequencies of these objects are compared with those in 
the corresponding portion of NGC 2403. Lacking $r-$band photometry for NGC 247, 
the specific frequencies were computed using $K-$band surface 
brightness measurements from Jarrett et al. (2003). 
The NGC 247 data have been shifted in brightness to match the distance modulus of 
NGC 2403, assuming a distance modulus of 27.9 for NGC 247, as computed 
from RGB-tip stars by Davidge (2006a). The impact on the NGC 247 data of adopting a 
distance modulus that is 0.5 magnitudes lower (i.e. $\mu_0 = 27.4$) is indicated by the 
arrow in each panel.

	It can be seen from Figure 17 that the specific frequencies 
of blue and red stars in NGC 247 are systematically 
lower than those in NGC 2403. The difference is largest among the blue stars, 
where the density of objects per unit $K-$band brightness 
is $0.4 - 0.6$ dex lower in NGC 247 than in NGC 2403. The 
differences between the two galaxies are smaller when RSGs are considered, 
falling in the range $0 - 0.4$ dex in the lower panel of Figure 17. That the specific 
frequencies of RSGs are in better agreement than that of massive main sequence 
stars suggests that any difference between the SFRs of NGC 247 and NGC 2403 was 
smaller a few tens of millions of years in the past.

	The comparisons in Figure 17 indicate that the density of young stars per 
unit stellar disk mass is higher in NGC 2403 than in NGC 247, and this is consistent 
with a higher recent SFR per unit stellar disk mass in NGC 2403. Such a result is 
consistent with the relative total FIR fluxes of these 
galaxies, which is an indicator of the recent SFR. Rice et al. (1988) give total FIR 
fluxes for both galaxies computed from IRAS observations. After correcting the NGC 247 
flux to the distance of NGC 2403 and then scaling the NGC 247 flux upwards by 0.8 
magnitudes to account for the difference in M$_K$ between the two galaxies, 
then the FIR flux in NGC 247 is roughly half that in NGC 2403, suggesting that 
the density of young stars should differ by $\sim 0.3$ dex. This is comparable to 
the difference between the NGC 247 and NGC 2403 curves in Figure 17. 

\subsection{An Extended Outer Disk in NGC 2403}

	A robust result of this study is that bright main sequence stars are 
traced out to R$_{GC} \sim 16$ kpc in the plane of the disk of NGC 2403. 
The scale length of the NGC 2403 disk is $\sim 2.1$ kpc (Kent 1987), 
and so the young stellar disk extends out to at least $\sim 7 - 
8$ scale lengths. This is consistent with other recent studies 
that have found that the stellar disks in nearby Sc galaxies 
typically extend out to 6 or more scale lengths. 
For example, Bland-Hawthorn et al. (2005) trace disk stars 
out to 14 kpc, or 10 scale lengths, in the Sc galaxy NGC 300, while 
Davidge (2003) and Tiede, Sarajedini, \& Barker (2004) 
argue that the disk of M33 extends to at least R$_{GC} \sim 12$ kpc, 
which corresponds to a distance in excess of 6 scale lengths.
Davidge (2006a) trace main sequence stars in NGC 247 out to 18 kpc, or $\sim 7$ 
scale lengths. The brightest stars in the outer disk of NGC 247 and NGC 
2403 have ages that are considerably less than the disk crossing time, 
suggesting that they formed {\it in situ}, and did not migrate into these regions. 
Signs of disk truncation are not seen in any of these studies, 
and so the disk sizes obtained from these data are lower limits. 

	Clues into the origin of the material from which the stars 
in the outer disk of NGC 2403 formed can be gleaned from studies of the stellar content. 
If the stars in the outer disk of NGC 2403 formed from gas 
that was accreted during early epochs, as might be expected from the Bullock 
\& Johnson (2005) simulations, then it is likely that 
there would be a large population of old stars in the outer disk, which formed during 
episodes of star formation that span the time since the material was accreted. For 
comparison, if material that has only recently been accreted has an 
angular momentum that is higher than that of other material in the disk 
then there should not be an underlying substrate of very old stars in the outermost 
regions of the disk. Estimates of the mean age in the outer disk might then provide 
insight into the epoch of gas accretion. 

	The outer regions of NGC 247 and NGC 2403 have different relative ages when 
compared with the main body of their disks. Whereas bright main sequence stars 
with ages $\leq 10$ Myr are found in the outermost regions of the disk of NGC 2403, 
the youngest stars in the outermost portions of the NGC 247 disk have ages $\geq 
40$ Myr (Davidge 2006b). In addition, the specific frequency of main sequence stars 
changes with radius in NGC 247, in the sense that main sequence stars contribute a higher 
fraction of the light in the outer disk than the inner disk (Davidge 2006b). Such 
a trend suggests that the outer disk has a younger photometrically-weighted age 
than the inner disk, as would be expected if the disk has grown from the inside out. 
For comparison, the specific frequency of main sequence stars in 
the outer disk of NGC 2403 appears to be lower than at intermediate radii, 
although the specific frequency measurements of RSGs suggests that this is likely a 
short-lived ($\leq 10$ Myr) phenomenon, and that the photometrically weighted age 
does not change with radius. A potential source of uncertainty in the NGC 
2403 specific frequency measurements is that they rely on the $r-$band photometry that 
was `extended by hand' by Kent (1987) to large radii. This being said, 
a highly fortuitous error is required to produce specific frequency measurements that 
do not change with radius in the RSG data.

	The chemical content of stars in the outer disk provides further insight 
into the origin of the material from which they formed. The narrow RSG plume in the CMD 
of NGC 2403 between R$_{GC} =$ 6 and 12 kpc suggests that these stars formed from 
material that had a metallicity that was spatially uniform over distances of many kpc. 
Abadi et al. (2003) argue that disks are assembled in large part from disrupted 
satellites, which presumably have diverse chemical enrichment histories. Indeed, 
in their investigation of the extended disk of M31, Ibata et al. (2005) argue that 
chemical homogeneity such as that seen among RSGs in NGC 2403 is not consistent with 
the outer disk originating from the accretion of numerous structures 
that experienced independent chemical enrichment paths prior to accretion, although 
Yoachim \& Dalcanton (2006) argue that such homogeneity might be expected if the 
majority of stars tend to come selectively from the largest satellites. 
Both arguments apply only to dry mergers, or to very recent gas-rich mergers. Indeed, 
if any structures accreted by NGC 2403 are gas dominated and were assimilated 
at an epoch that is much older than the crossing time of the disk, then the current 
generation of stars could form from chemically homogeneous material, 
as there will then be ample time for the gas to mix.

\subsection{AGB Stars in the Extraplanar Regions of NGC 2403}

	The WIRCam data have been used to investigate the stellar content on the minor 
axis of NGC 2403 at a point where disk contamination is negigible, and 
a modest population of sources that are likely AGB stars is found. 
With a peak brightness M$_K \sim -8$, isochrones suggest that these 
stars may belong to a population with an age of a few Gyr.
However, there are uncertainties in this age estimate. 
AGB stars are highly evolved, and there are uncertainties in the physics used 
to create the models that form the basis for the age estimate. In addition, a core 
assumption is that the stars are metal-poor. This assumption seems secure, as Davidge 
(2003) found that RGB stars at the location of the NE field have [Fe/H] $\sim -2.2$.

	Another source of uncertainty in the age estimate is that a 
large fraction of AGB stars are photometric variables, and these 
variations smear the true AGB-tip brightness. Observations of LPVs in 
the Galaxy (e.g. Glass et al. 1995) and nearby galaxies (e.g. Davidge \& Rigaut 2004) 
indicate that the amplitude of the photometric variations in $K$ are typically $\pm 0.5$ 
magnitudes. While variability will introduce systematic errors in age estimates, it may 
be possible to account for this using multi-epoch observations of the same field. Even 
then, when observing objects near the faint limit of a data set there is a bias to detect 
only those variables that are near the peak of their light curves, and this causes the 
AGB-tip brightness to be systematically overestimated and the age underestimated.

	NGC 2403 is not unique among late-type disk galaxies 
in containing a population of intrinsically bright extraplanar AGB stars. 
Davidge (2006a) found a population of luminous AGB stars at a projected 
distance of 12 kpc above the disk of NGC 247. 
The peak $i'$ brightness of these stars is consistent with an age $\sim 3$ Gyr, 
although this age estimate is prone to the same uncertainties discussed previously.

	Seth et al. (2005) find an extraplanar component in edge-on 
disk systems that is moderately metal-poor and spans a range of 
ages, with the youngest stars closest to the disk and the oldest stars furthest 
from the disk. These stars likely belong to the thick disk. 
Seth et al. (2005) argue that at least some of these stars may have formed 
in the thin disk, but were scattered out of the disk plane by dynamical interactions. 
Yoachim \& Dalcanton (2006) argue that the bulk of the stars were probably formed in 
pregalactic fragments, that were shredded during the early stages of disk formation.
The bright AGB stars found in NGC 2403 and NGC 247 are located much further off of the 
disk plane than expected for thick disk stars; for example, in the Milky-Way the thick 
disk has a scale height of $\sim 1 - 1.5$ kpc (e.g. Gilmore, Wyse, \& Kuijken 1989). 

	Mouhcine et al. (2005) probed the extraplanar regions of 
nearby spiral galaxies; NGC 2403 was not in their sample. They targeted 
fields on or near the minor axis, and sample extraplanar distances 
similar to those probed in NGC 2403 with WIRCam. 
One conclusion from the Mouhcine et al. (2005) study is that there is no evidence 
for an intermediate age population in their CMDs. However, 
the CMDs of NGC 253, NGC 3031, NGC 4258, and NGC 4945 shown in their Figures 2 and 3 have 
a substantial spray of stars that extend well above the RGB-tip in $I$; in the case of 
the last two galaxies the AGB appears to extend a magnitude or more above the RGB-tip. 
The presence of an extended AGB above the RGB-tip is not an ironclad signature of an 
intermediate age population, especially in systems that are metal-rich. 
Still, the colors of RGB stars in these systems suggest that [Fe/H] $< -1$, and the 
models of Girardi et al. (2002) suggest that at least some of the galaxies in the 
Mouchine et al. (2005) sample may harbor an extraplanar component with an age $< 10$ Gyr. 

	We close this part of the discussion by noting that NGC 247 
is in the sample of galaxies observed by Mouhcine et al. (2005), and their CMD of 
this galaxy shows no stars above the RGB-tip. While this is at odds with the conclusions 
reached by Davidge (2006a), Mouhcine et al. (2005) adopted a distance modulus that is 
0.6 magnitudes lower than that measured from the RGB-tip by Davidge (2006a). 
If the RGB-tip distance modulus of Davidge (2006a) is adopted for NGC 247 then a 
population of AGB stars above the RGB-tip is present in the Mouchine et al. CMD. 

\acknowledgements{It is a pleasure to thank the anonymous referee for comments that 
greatly improved the manuscript.}

\clearpage

\begin{table*}
\begin{center}
\begin{tabular}{cccccc}
\tableline\tableline
Name & $K$ & $J-K$ & $r'-K$ & $V-K$ & r (pc) \\
\tableline
F1 & 16.78 & 0.64 & -- & -- & 5.0 \\
F14 & 17.90 & 1.51 & 5.44 & 6.34 & 4.0 \\
F16 & 16.44 & 0.63 & 2.54 & 2.89 & 2.7 \\
F19 & 17.94 & 1.80 & 4.61 & 5.36 & 8.6 \\
F46 & 18.82 & 0.42 & 2.01 & 2.27 & 10.4 \\
\tableline
\end{tabular}
\end{center}
\caption{Photometry of Battistini et al. (1984) Cluster Candidates}
\end{table*}

\clearpage

\begin{table*}
\begin{center}
\begin{tabular}{ccccc}
\tableline\tableline
RA & Dec & $K$ & $J-K$ & r (pc) \\
\tableline
07:38:16.8 & 65:42:40.5 & 17.94 & 0.90 & 4.5 \\
07:37:27.0 & 65:28:20.5 & 15.98 & 0.26 & 1.8 \\
07:37:21.5 & 65:28:16.6 & 15.84 & 0.63 & 1.4 \\
07:36:17.3 & 65:29:00.5 & 17.95 & 0.99 & 14.0 \\
07:36:04.4 & 65:34:00.3 & 15.79 & 0.67 & 3.6 \\
07:35:05.9 & 65:45:27.5 & 16.70 & 0.66 & 3.6 \\
\tableline
\end{tabular}
\end{center}
\caption{New Globular Cluster Candidates}
\end{table*}

\begin{figure}
\figurenum{1}
\epsscale{0.75}
\plotone{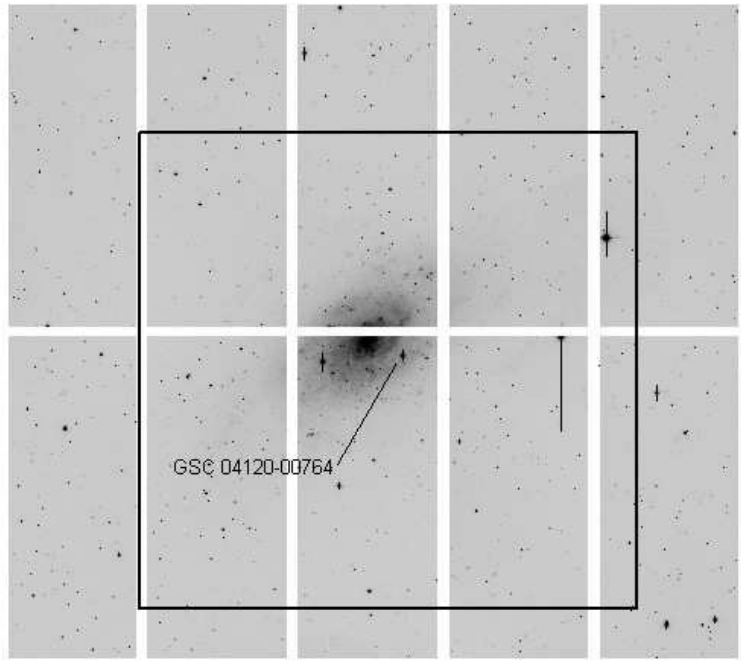}
\caption
{The final $i'$ images of the ten MegaCam CCDs that sample the disk and inner halo of 
NGC 2403. The box marks the WIRCam field, while the bright star that is the 
guide source for the CFHT AO observations is labelled.}
\end{figure}

\clearpage

\begin{figure}
\figurenum{2}
\epsscale{0.75}
\plotone{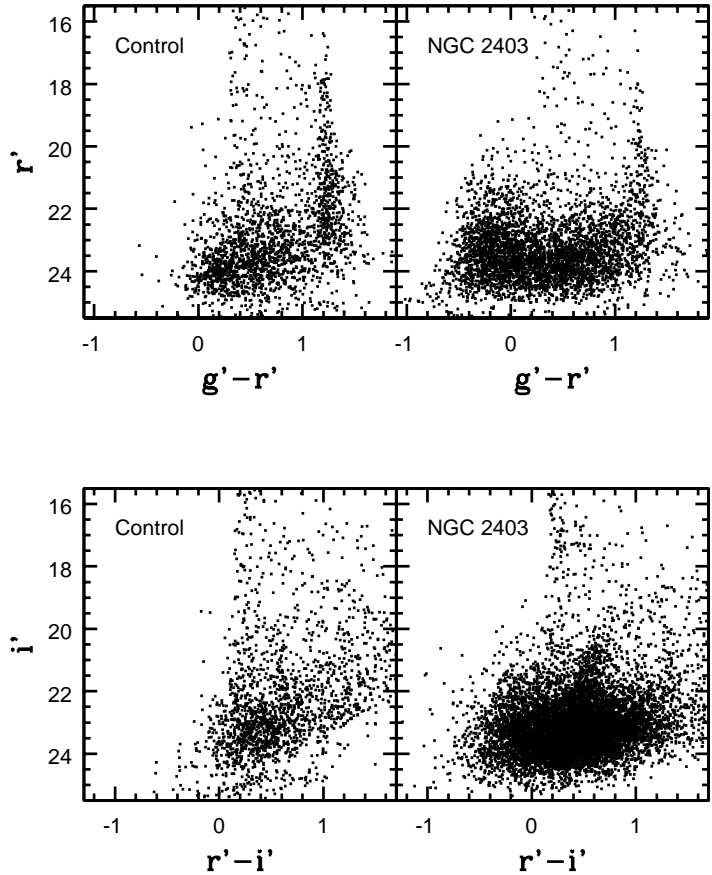}
\caption
{The $(r', g'-r')$ and $(i', r'-i')$ CMDs of selected MegaCam CCDs. The left hand 
column shows the CMDs of sources in CCDs 12 and 25, which sample areas on the 
sky where the density of sources belonging to NGC 2403 is negligible; 
consequently, these CCDs constitute a control field for monitoring contamination 
from foreground stars and background galaxies. The CMDs in the right hand column show 
sources in CCDs 15 and 22, which sample the disk of NGC 2403 at intermediate 
distances. See the text for a discussion of the various features in the CMDs.}
\end{figure}

\clearpage

\begin{figure}
\figurenum{3}
\epsscale{0.75}
\plotone{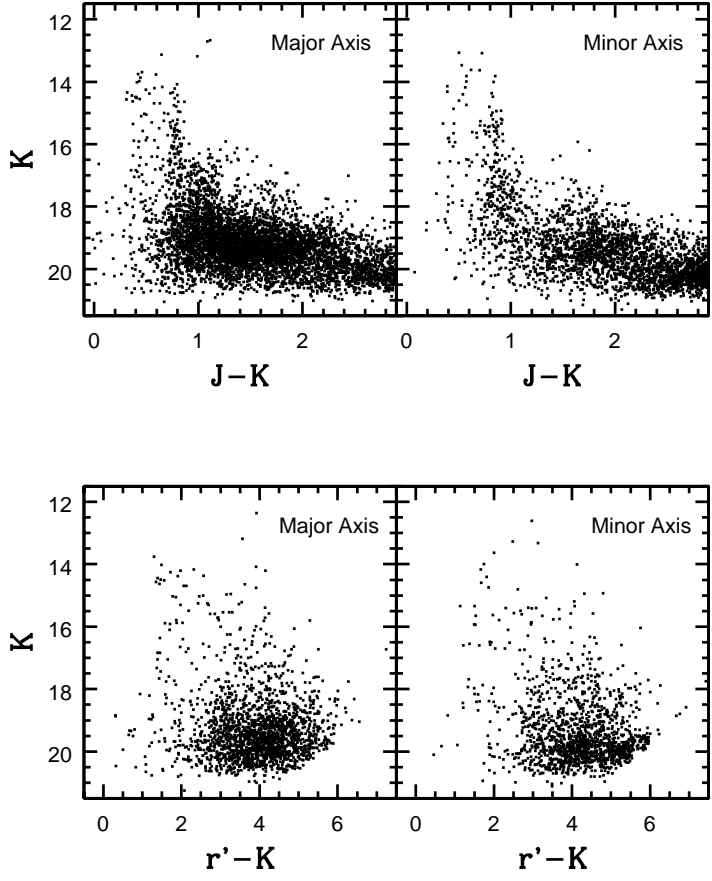}
\caption
{The $(K, J-K)$ and $(K, r'-K)$ CMDs of sources in two quadrants of the  WIRCam array, 
that sample the major and minor axes of NGC 2403. 
Foreground stars have $J-K \leq 1.0$ and $r'-K \leq 2.5$, 
while many of the reddest objects are background galaxies. 
RSGs have $K > 16$, with $J-K \sim 1$ and $r'-K \sim 3$.
The concentration of objects with $K > 18.5$ and colors $J-K < 2.2$ and 
$r'-K > 3$ is dominated by stars evolving on the AGB.}
\end{figure}

\clearpage

\begin{figure}
\figurenum{4}
\epsscale{0.75}
\plotone{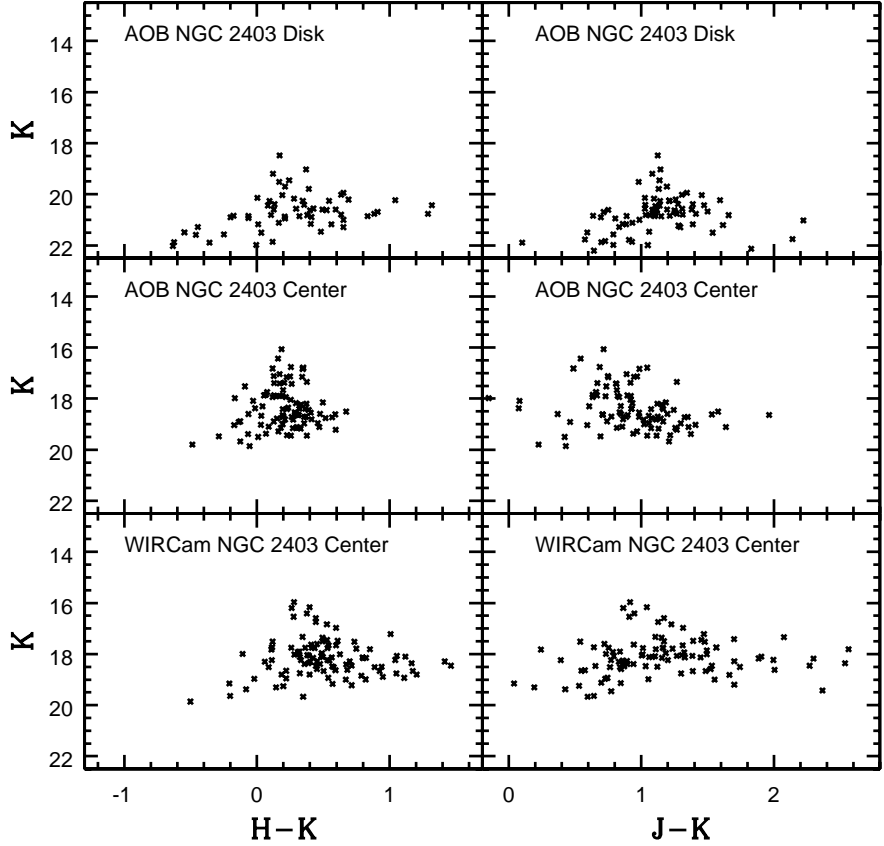}
\caption
{The $(K, H-K)$ and $(K, J-K)$ CMDs obtained from 
observations recorded with the CFHT AOB (top and middle rows) and WIRCam (bottom row). 
Despite uncertainties in the photometric calibration, the peak AGB 
brightness in the disk field is consistent with what 
is seen in the WIRCam CMDs of the disk. Note the reasonable agreement between the 
CMDs of the center field as observed with the AOB and as extracted from the 
WIRCam data. This agreement suggests that crowding does not affect the WIRCam 
observations of bright stars in NGC 2403.}
\end{figure}

\clearpage

\begin{figure}
\figurenum{5}
\epsscale{0.75}
\plotone{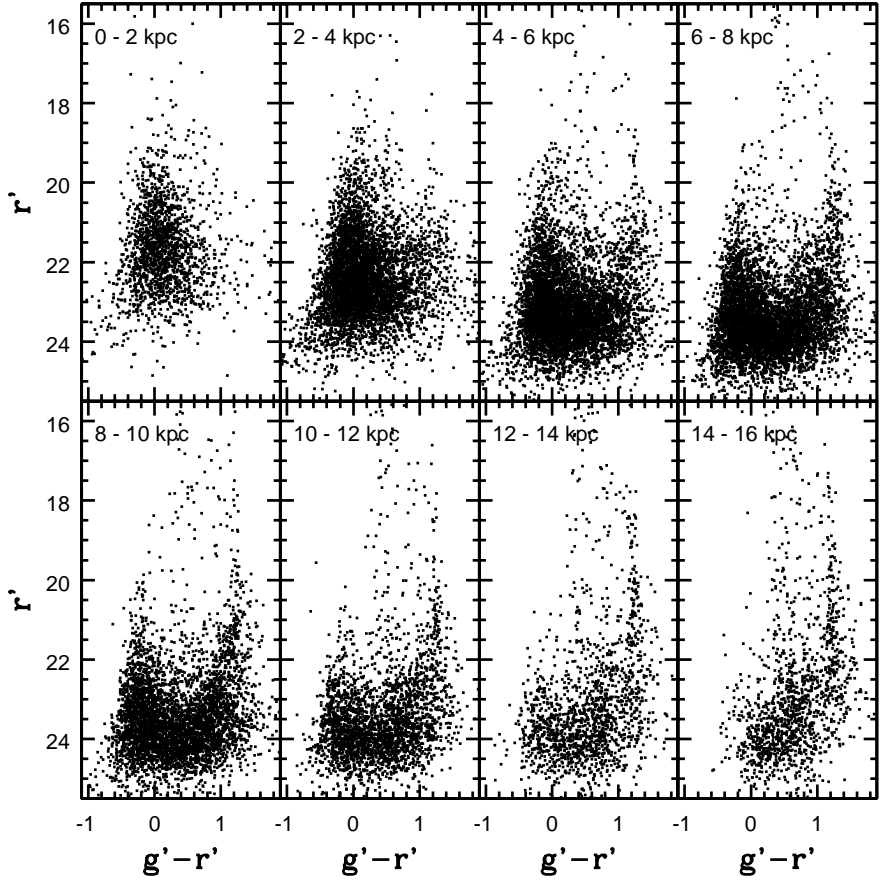}
\caption
{The $(r', g'-r')$ CMDs of stars at different galactocentric radii 
in the disk of NGC 2403. The CMDs for the $0 - 2$ and $2 - 4$ kpc intervals 
do not go as deep as at larger distances because of the high stellar densities 
near the center of NGC 2403. The foreground star sequences 
become more pronounced towards progressively larger R$_{GC}$ because of the 
larger areas on the sky that are sampled in each radial interval.
The main sequence is seen in the CMD of the $12 - 14$ kpc interval, indicating that 
(1) the disk of NGC 2403 extends out to at least $\sim 7$ scale lengths, and (2) 
recent star formation has occured in the outer disk of NGC 2403.} 
\end{figure}

\clearpage

\begin{figure}
\figurenum{6}
\epsscale{0.75}
\plotone{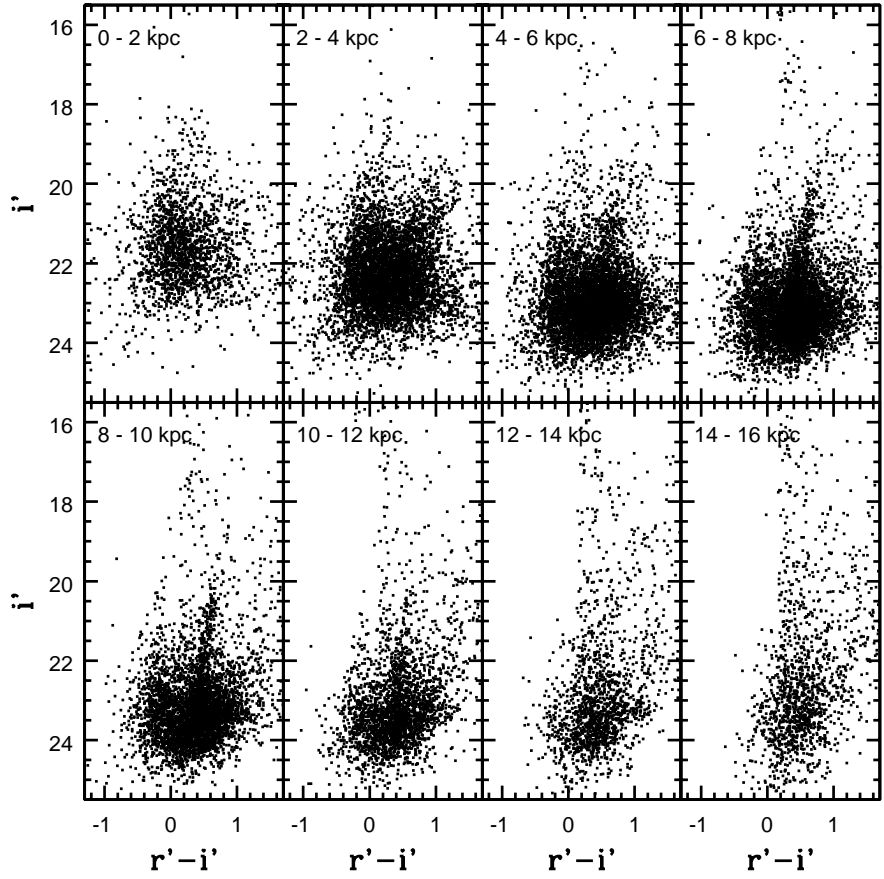}
\caption
{The same as Figure 5, but showing the $(i', r'-i')$ CMDs. Note that RSGs form a 
well-defined finger on the CMDs of objects with R$_{GC} > 6$ kpc}
\end{figure}

\clearpage

\begin{figure}
\figurenum{7}
\epsscale{0.75}
\plotone{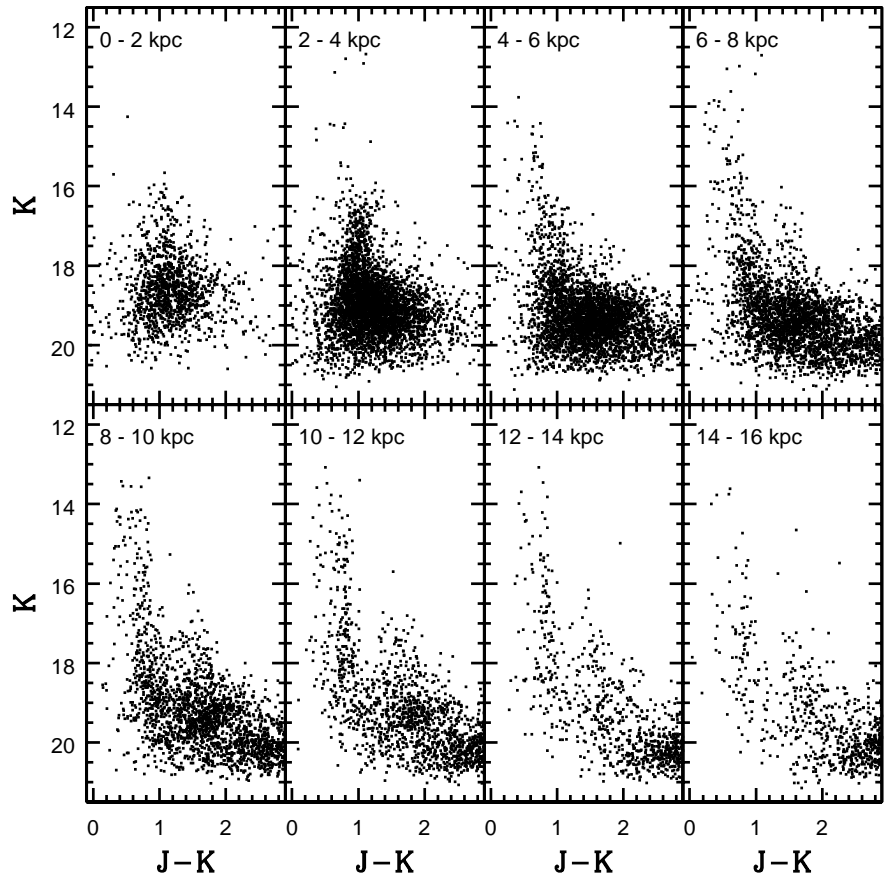}
\caption
{The same as Figure 5, but showing the $(K, J-K)$ CMDs.}
\end{figure}

\clearpage

\begin{figure}
\figurenum{8}
\epsscale{0.75}
\plotone{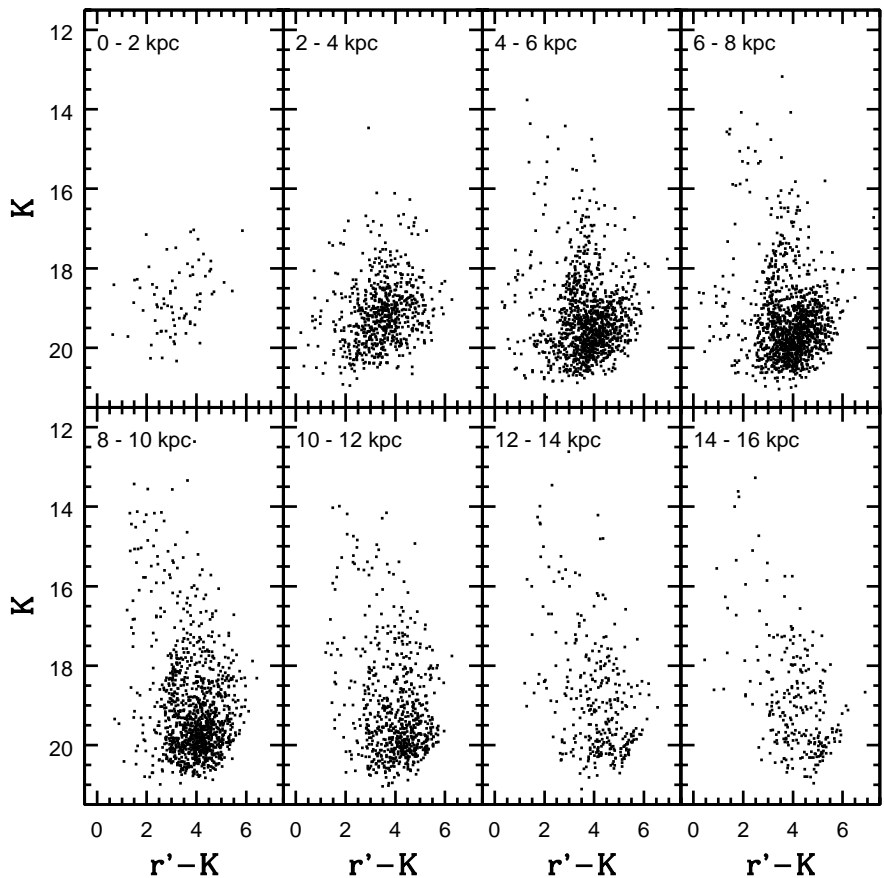}
\caption
{The same as Figure 5, but showing the $(K, r'-K)$ CMDs. 
Only a modest number of stars are in this data set, due to the 
faint limit of the $r'$ data, and the loss of sky coverage arising from 
gaps between arrays in both detector mosaics. Still, there is a pronounced 
AGB sequence with $K > 18.5$ in the CMDs of objects with R$_{GC}$ between 4 and 10 kpc.}
\end{figure}

\clearpage

\begin{figure}
\figurenum{9}
\epsscale{0.75}
\plotone{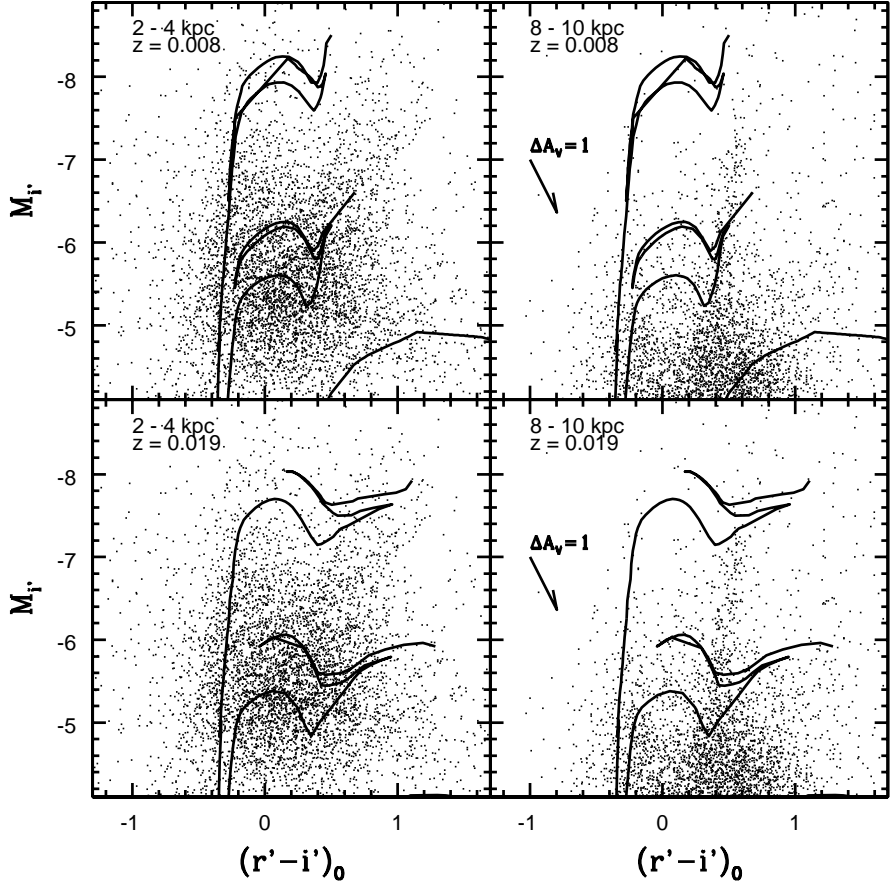}
\caption
{The $(M_{i'}, (r'-i')_0)$ CMDs of stars with R$_{GC}$ between 2 and 4 kpc (left hand 
column) and 8 and 10 kpc (right hand column). The three sequences plotted in each 
panel are isochrones from Girardi et al. (2004) that have log(t$_{yr}$) = 7.0, 
7.5, and 8.5. Models with Z = 0.008 (top row) and Z = 0.019 (bottom row) are 
shown. An internal extinction of A$_V = 0.15$ magnitudes (Pierce \& Tully 1992) 
has been applied to the $8 - 10$ kpc data. An additional 
internal extinction of A$_V = 0.3$ magnitudes has been 
applied to the $2 - 4$ kpc data, based on the color of the main sequence 
in this portion of the galaxy (\S 4.1). A reddening vector, with a length corresponding 
to A$_V = 1$ mag, is shown in the right hand panels. While the Z = 0.008 models match 
the RSG plume in the $8 - 10$ kpc CMD, there are RSGs in the $2 - 4$ kpc interval 
with colors that are consistent with Z $\sim 0.019$.}
\end{figure}

\clearpage

\begin{figure}
\figurenum{10}
\epsscale{0.75}
\plotone{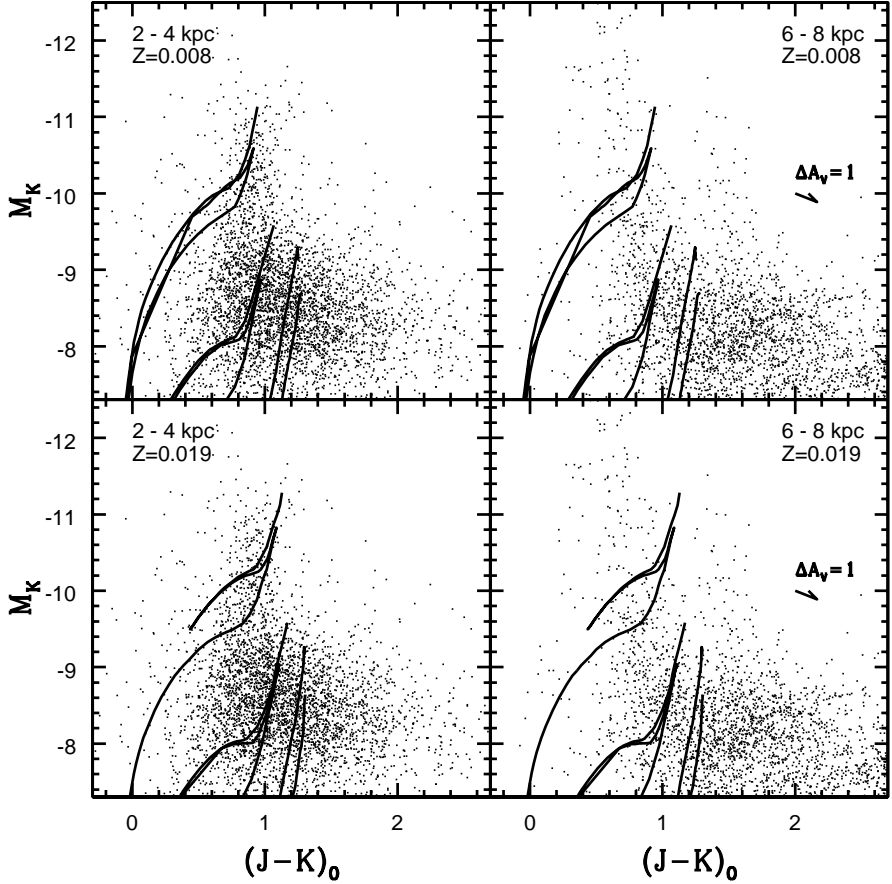}
\caption
{The $(M_{K}, (J-K)_0)$ CMDs of stars with R$_{GC}$ between 2 and 4 kpc (left hand 
column) and 6 and 8 kpc (right hand column). 
An internal extinction of A$_V = 0.15$ magnitudes (Pierce \& Tully 1992) 
has been applied to the $6 - 8$ kpc data. An additional 
internal extinction of A$_V = 0.3$ magnitudes has been 
applied to the $2 - 4$ kpc data, based on the color of the main sequence in this 
portion of the galaxy (\S 4.1). A reddening vector, with a length corresponding to 
A$_V = 1$ mag, is shown in the right hand panels. Isochrones with ages log(t$_{yr}$) = 
7.0, 7.5, 8.5, and 9.0 and Z = 0.008 (top row) and Z = 0.019 (lower row) from Girardi 
et al. (2002) are also shown. Note that the red loci of the Z = 0.008 models and 
Z = 0.019 models differ by only $\sim 0.1$ magnitudes, and that the foreground star 
sequence in the $6 - 8$ kpc interval is much more prominent than in the 
$2 - 4$ kpc interval, complicating efforts to probe the RSG population in this radial 
interval.}
\end{figure}

\clearpage

\begin{figure}
\figurenum{11}
\epsscale{0.75}
\plotone{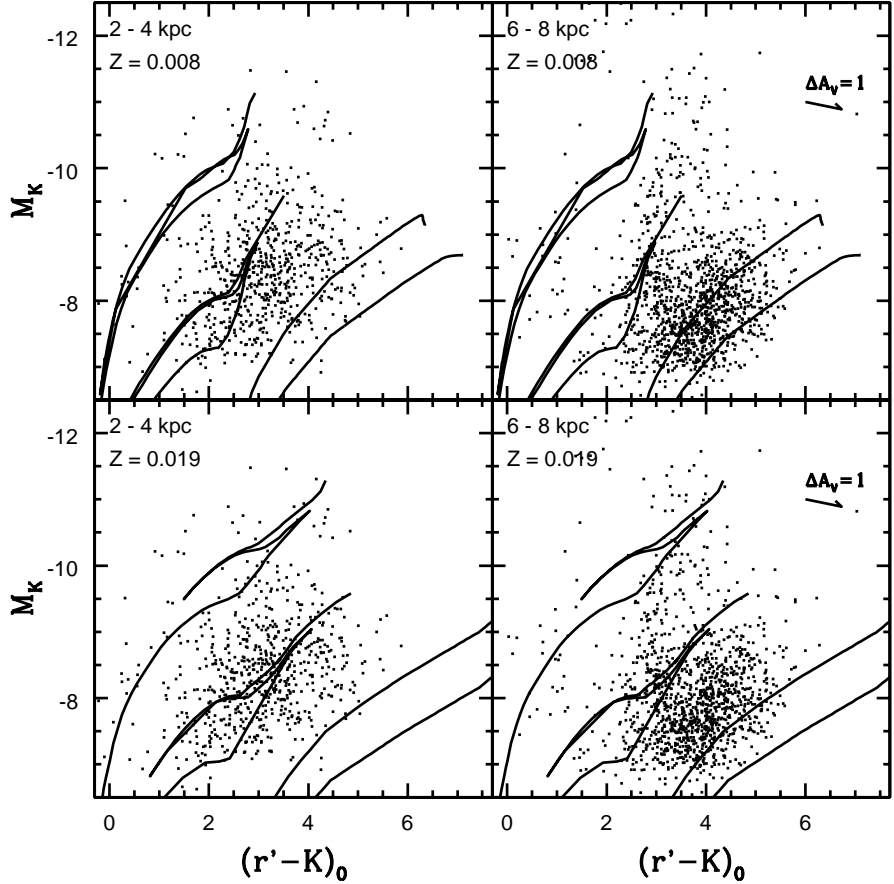}
\caption
{The same as Figure 9, but showing the $(M_{K}, (r'-K)_0)$ CMDs. RSGs in the $2 - 
4$ kpc CMD occupy a range of colors, as was the case in the $(i', r'-i')$ CMD, 
and the Z=0.008 and Z = 0.019 models roughly bracket the colors of stars in this radial 
interval. The red locus of the Z=0.008 models better matches the plume of 
stars with $r'-K \sim 3$ in the $6 - 8$ kpc CMD than the Z=0.019 models. 
These data support (1) a spread in the metallicity of RSGs in the 
inner regions of NGC 2403, and (2) a uniform RSG metallicity when R$_{GC} > 6$ kpc.} 
\end{figure}

\clearpage

\begin{figure}
\figurenum{12}
\epsscale{0.75}
\plotone{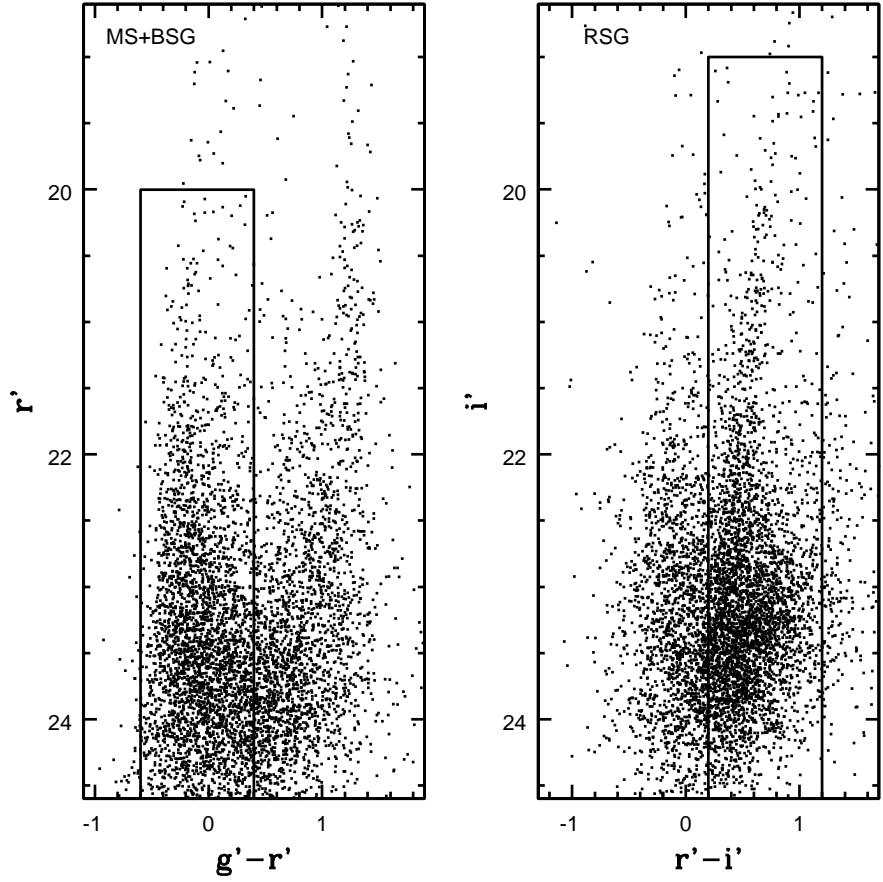}
\caption
{The $(r', g'-r')$ and $(i', r'-i')$ CMDs of stars with R$_{GC}$ 
between 6 and 8 kpc. The boundaries used to identify 
main sequence $+$ BSG stars (left hand panel) and RSGs (right hand 
panel) are indicated.}
\end{figure}

\clearpage

\begin{figure}
\figurenum{13}
\epsscale{0.75}
\plotone{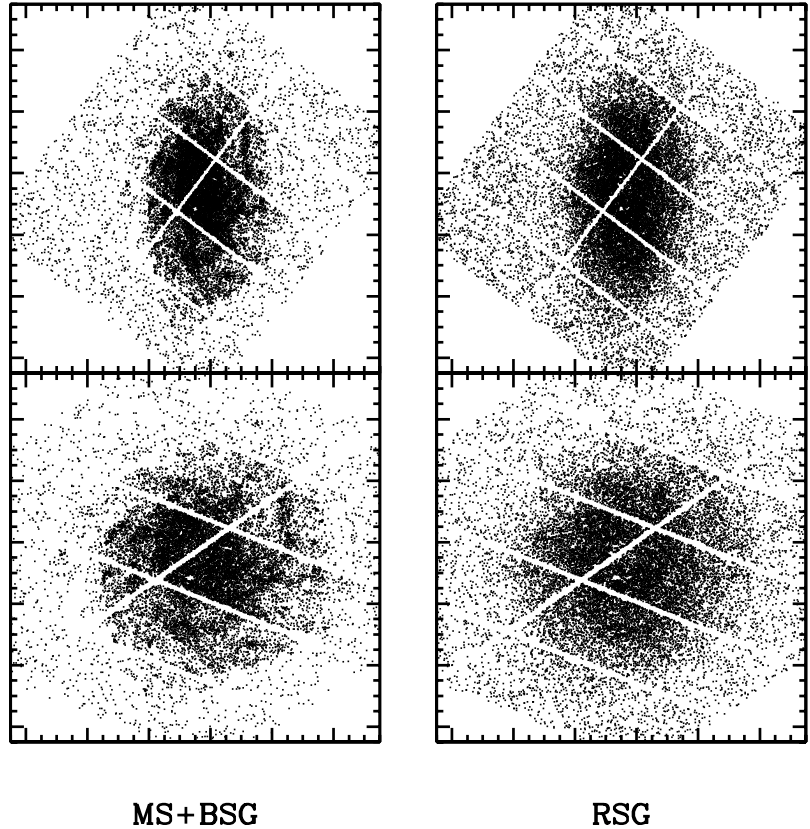}
\caption
{The observed (top row) and de-projected (lower row) distributions of 
main sequence stars and BSGs and RSGs, as identified using the boundaries in 
Figure 12. The spatial co-ordinates have been rotated so 
that the major axis of NGC 2403 runs parallel to the vertical axis. 
The distribution of RSGs is more diffuse than that of the MS$+$BSG stars, 
which track the spiral structure in NGC 2403. The absence of isolated star-forming 
regions in the RSG distribution suggests that these stars have a velocity 
dispersion $\sim \pm 10$ km sec$^{-1}$ if they have a typical age of 50 million years.}
\end{figure}

\clearpage

\begin{figure}
\figurenum{14}
\epsscale{0.75}
\plotone{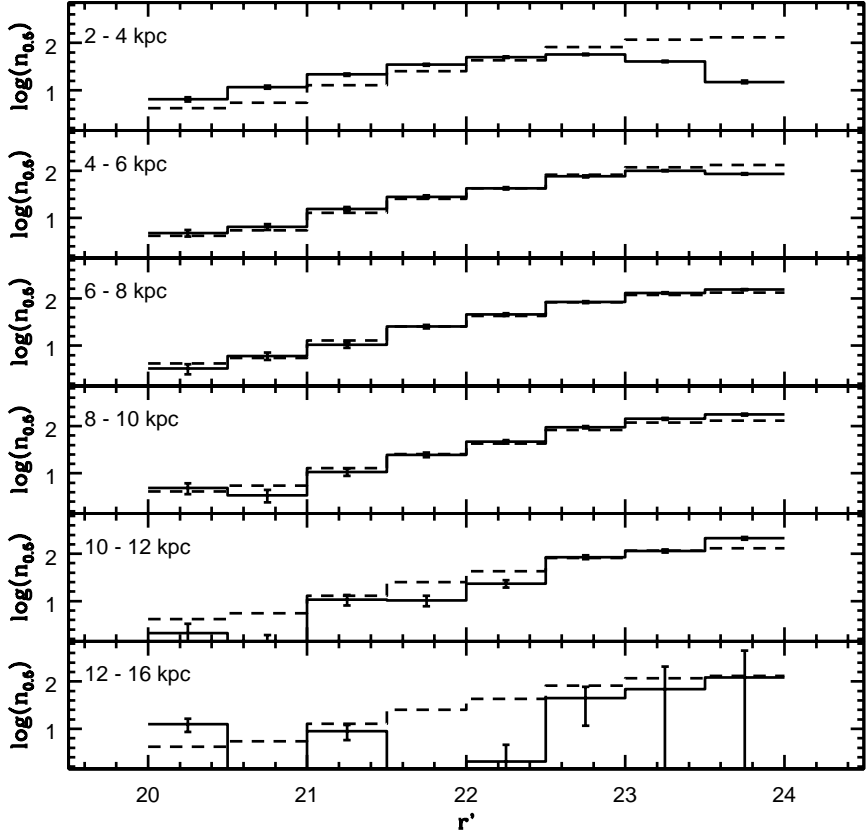}
\caption
{The specific frequency of main sequence stars and BSGs, as defined 
by the boundaries in the left hand panel of Figure 12. log(n$_{0.5}$) 
is the number of stars per 0.5 magnitude interval in $r'$, scaled to a system with 
M$_{r} = -15$ and corrected for background and foreground sources. The dashed line is 
the mean specific frequency of main sequence stars and BSGs between 4 and 12 kpc from the 
center of NGC 2403. Note (1) the elevated number of main sequence stars near the bright 
end in the $2 - 4$ kpc interval, and (2) the departures from the mean curve when 
R$_{GC} >$ 10 kpc.}
\end{figure}

\clearpage

\begin{figure}
\figurenum{15}
\epsscale{0.75}
\plotone{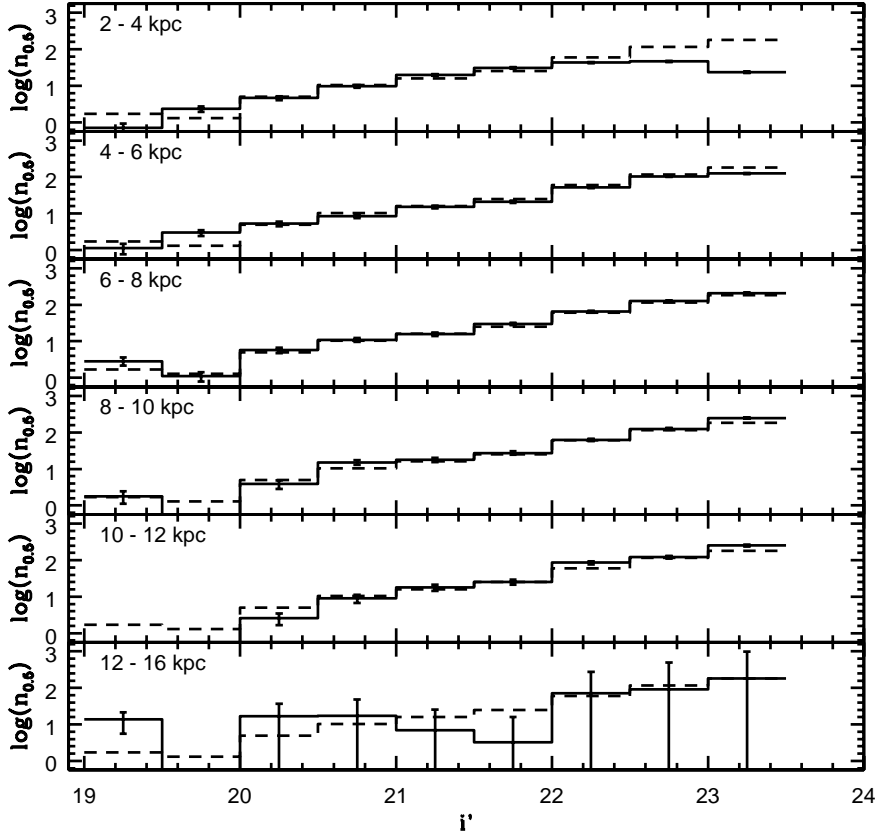}
\caption
{The same as Figure 14, but showing the specific frequency of stars with 
the photometric properties of RSGs, as defined in the right hand panel of Figure 12. 
log(n$_{0.5}$) is the number of stars per 0.5 magnitude interval in $i'$, scaled to 
an integrated system brightness M$_{r} = -15$ and corrected for background and foreground 
sources. The specific frequency of RSGs throughout the disk of NGC 2403 is 
constant when $i' > 20$, indicating that these stars follow the integrated 
$r-$band light profile of NGC 2403.}
\end{figure}

\clearpage

\begin{figure}
\figurenum{16}
\epsscale{0.75}
\plotone{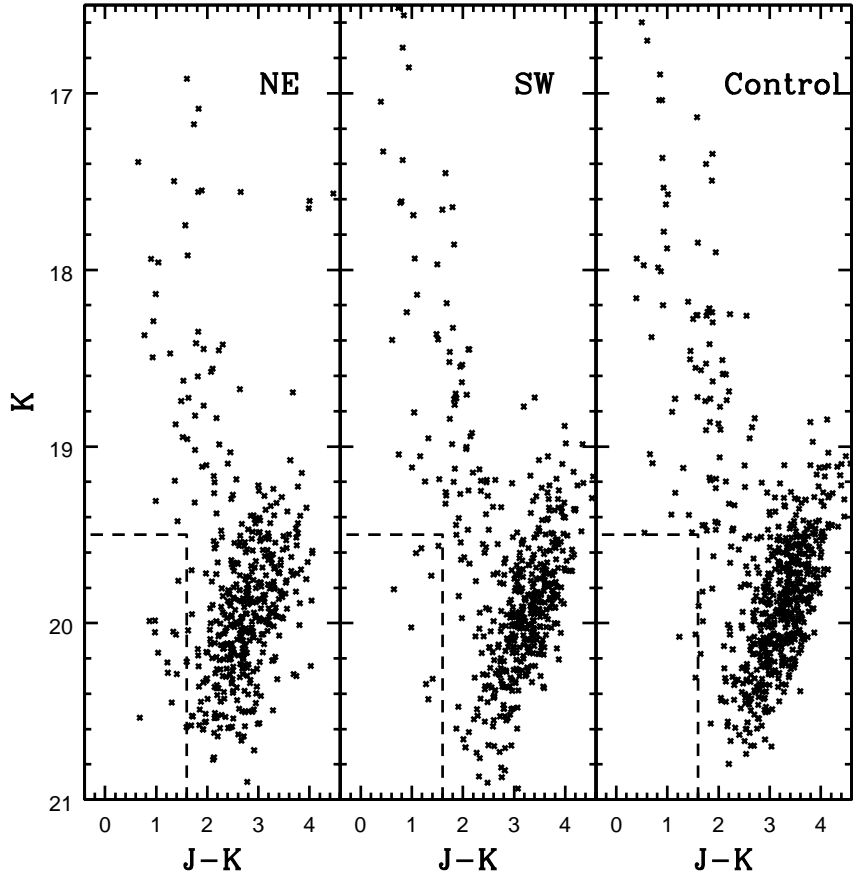}
\caption
{The $(K, J-K)$ CMDs of the NE and SW fields, which sample the extraplanar regions 
along the minor axis of NGC 2403. The CMD of a 
control field is also shown. Note that the three CMDs are very similar, except in the 
region marked with dashed lines, which is where M giants in NGC 2403 are expected.}
\end{figure}

\clearpage

\begin{figure}
\figurenum{17}
\epsscale{0.75}
\plotone{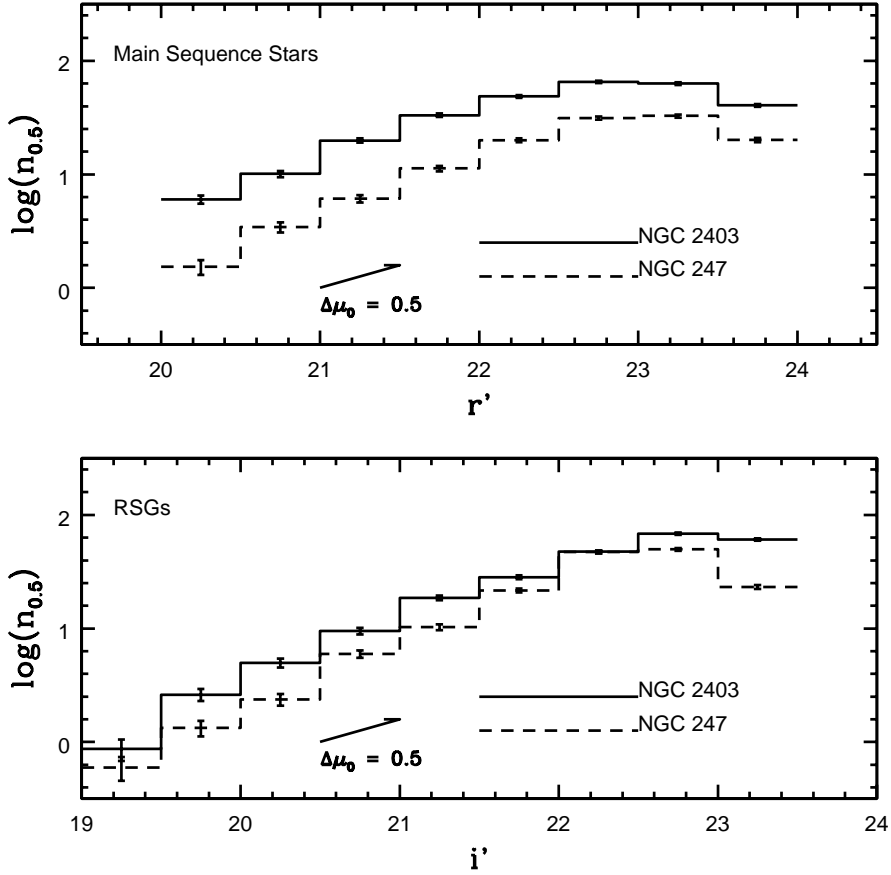}
\caption
{The specific frequencies of MS$+$BSGs (top panel) and RSGs (lower panel) in NGC 247 
and NGC 2403. N$_{0.5}$ is the number of stars per 0.5 magnitude interval that would 
appear in a system with M$_K = -16$, after scaling the NGC 247 
and NGC 2403 LFs based on the $K-$band surface photometry of 
Jarrett et al. (2003). The LFs were constructed from stars with R$_{GC}$ between 2 
and 6 kpc in each galaxy. The NGC 247 data have been shifted along the horizontal axis 
by 0.4 magnitudes to account for the difference in distance between the two 
galaxies. The error bars show the uncertainties due to counting statistics, while 
the arrows indicate the approximate impact on the NGC 247 data of adopting a distance 
modulus of 27.4 for that galaxy, which is 0.5 mag smaller than the baseline 
distance modulus adopted for the comparison. It is evident that the use of the smaller 
distance modulus would not significantly alter the main conclusion of this comparison, 
which is that the number density of blue and red stars in NGC 247 
is lower than in NGC 2403. This suggests that the SFR during the past $\sim 10$ Myr 
was $0.4 - 0.6$ dex lower in the former than in the 
latter. That there is better agreement amongst the red stars suggests that the SFRs in 
the two galaxies were in better agreement a few tens of millions of years in the past.}
\end{figure}

\end{document}